%% file: main.tex
\documentclass{iopjournal}

\usepackage{amsmath}
\usepackage{amssymb}
\usepackage{url}
\usepackage{cite}

\makeatletter
\renewcommand{\articletype}[1]{{\vspace*{-8mm}\noindent \Large \sffamily}

\vspace*{8mm} \noindent\reversemarginpar
\marginpar{\vspace{-3mm} {\color{gray}\hrule} \ \\ Crossmark\\ {\color{gray}\hrule} \ \\ \tiny {\sffamily RECEIVED} {\small \\ dd Month yyyy}\\ \\ {\sffamily REVISED} {\small \\ dd Month yyyy}}{\scriptsize \sffamily{\bfseries \MakeUppercase{#1}}}}
\makeatother
\begin{document}

% Main text is the color-feature SDSS column. Old text: \sdsscolorfalse in sdss_color_switch.tex.
\input{sdss_color_switch}

\articletype{Paper}

\title{A Radiation-Hardened, Fault-Aware Quantized Neural Network That Self-Heals}

\author{Peter Xiangyuan Ma$^{1,*}$\orcid{0000-0001-8975-3719} and Joshua S. Bloom$^{1}$\orcid{0000-0002-7777-216X}}

\affil{$^1$Department of Astronomy, University of California, Berkeley, CA 94720, USA}

\affil{$^*$Author to whom any correspondence should be addressed.}

\email{peter\_ma@berkeley.edu}


\begin{abstract}
Deep learning (DL) models are increasingly being deployed into radiation-challenging environments in spacecraft, particle accelerators, and, soon, space-based data centers. Instantiated as a learned orchestration of numbers stored in persistent memory, models are susceptible to continual degradation as high-energy particles corrupt data. At inference time, new data passed through the network fundamentally break the off-line assumptions and guarantees of a static model determined during training. To preserve the quality and efficacy of the models, fault tolerance against bit flips is a clear objective. In this work we developed a runtime method for repairing the quantized weights of neural networks (NNs) after bit flips on commodity, commercial off-the-shelf (COTS) hardware, without storing extra replicas. Our work extends the well-established Fault-Aware Training (FAT) paradigm by adding an alignment penalty to concentrate undamaged activations and weights into a low-rank subspace, and we then reconstruct flagged corrupted coordinates from the uncorrupted survivors. We call this method {\it Projection Healing} and apply it to six regression benchmarks on quantized NNs. Holding a FAT checkpoint fixed and toggling only the runtime heal gives a further ${\sim}23\%$ reduction in the native-unit excess integral. Across those datasets the method extends the operational scrub interval by a median of $2.04\times$ at a $+5\%$ RMSE degradation and $1.96\times$ at $+50\%$, compared with $1.62\times$ and $1.07\times$ for classical Single Error Correction, Double Error Detection (SECDED). This introduces a new path towards designing resilient data-driven models, complementing established hardware redundancy, ECC, and FAT-based techniques.
\end{abstract}

\section{Introduction}
Deep learning algorithms are now frequently deployed into radiation-challenging environments, including low-Earth orbit \cite{izzo2019machine, furano2020machine}, aerospace electronics \cite{normand1996single, ts62396_2006, taber1995single}, nuclear facilities \cite{lecorre2024chf}, and particle accelerators \cite{albertsson2018machine, radovic2018machine, guest2018deep}. These environments present a unique risk of introducing hardware-level faults due to incident particles ionizing a track through the silicon in the hardware. This results in flipped bits which, when the deposited charge exceeds a critical threshold of the transistor node, cause computation faults manifested at the software level as bit flips (if left uncorrected). The typical solution to this problem requires fabricating custom edge silicon where the designer introduces hardware redundancy to strategically protect critical components on a chip \cite{weng2026priorifi}. Examples of these custom-designed solutions are full triple modular redundancy (TMR) where, at the cost of $3\times$ chip footprint and redundant calculations \cite{lyons1962use, weng2024fkeras}, one can protect datapaths and computation. However, this often leaves the Static Random Access Memory (SRAM) underprotected as dense memory arrays scale unfavorably to this method in both footprint and power consumption, falling back to standard software error correcting codes (ECC). Commercial off-the-shelf (COTS) hardware used in small satellites often faces similar challenges where memory typically has no ECC \cite{satyanarayanan2017emergence}. This work targets that case with a software repair on the processor and memory a commodity part already has, rather than a radiation-hardened fabrication process or a custom datapath.

When extrapolated into the future, these risks are further elevated. As transistors shrink \cite{dodd2003basic, baumann2005radiation} and compute densifies, the probability of \textit{multiple} faults increases with future deployment. For example, conventional frontier deep learning accelerators, such as the Nvidia H100, use 3D-stacked high-bandwidth memory for model weights. That stacking raises the chance of spatially correlated bit flips. Stacked HBM itself predates the H100. This fundamentally shifts us from the realm of Single Event Upset (SEU) flips, where one cell is disrupted, to Multi-Bit Upset (MBU) flips, where a cluster of adjacent cells is affected \cite{black2008characterizing}. As the computing paradigm advances, we expect the next generation of deployed model weights to remain an unprotected radiation target under the status quo.

Needless to say, there are further engineering constraints to account for. In nearly all of the deployment settings, there is a non-negotiable need for systems to run on the edge with constraints on bandwidth, power, physical footprint, latency, and cost \cite{denby2020orbital, zhou2019edge}. In most settings, cheap, small, power-efficient commodity solutions are ideal, and in many categories, the current solution of custom-fabricated hardware redundancy underserves in many areas.

\section{Background and Related Work}
\label{sec:fault_tolerant_dnns}
Classically, fault tolerance uses a mix of spatial and algorithmic redundancy  \cite{koren2020fault}. As mentioned, TMR \cite{lyons1962use} triplicates blocks and votes, at over $3\times$ area and power, and ECC algorithms like Single Error Correction, Double Error Detection (SECDED) protect memory at a cost of storage. Other methods used commonly are checksums called Algorithm-Based Fault Tolerance \cite{huang1984algorithm} to verify computations. This idea has been adapted for Deep Neural Networks (DNNs) \cite{zhao2021ft, hari2020making}. The downside of this method is it adds memory overhead and also introduces unrecoverable threats as checksums can also be unprotected.

There are also a number of co-designed algorithms custom for DNNs specifically. For example, activation clipping is a common strategy to limit intermediate values from catastrophic failures due to floating-point exponent bit flips \cite{ma2024rednet, santos2024diehardnet, chen2020ranger, mousavi2024proact, ghavami2021fitact}. The downside is this manual clipping primarily protects classification models and leaves regression-based tasks underserved. This is because, unlike classifiers, regression models do not exhibit terminal-phase neural collapse \cite{papyan2020prevalence} and depend on \textit{both} the direction and the magnitude of latent vectors. Amplitude clipping used to harden classifiers \cite{qiao2019ftclipact} can be damaging, inert, or too rigid for regression. Furthermore, a quantization of the model at deployment time (such as to INT8) often alleviates most of the risks in these catastrophic failure modes by reducing the radiation sensitivity and dynamic cross-section of neural networks \cite{jacob2018quantization, libano2020understanding, libano2021how}, at a cost of absolute model performance. However, in the current state, most production models undergo quantization before deployment for inference throughput and hence act as a double protection against faults.

This leaves the remaining bits to carry more significance in the performance of the model. This changes the risk profile as it has been shown that single bit upsets are more likely to cause critical failures \cite{libano2020understanding, libano2021how} where small targeted attacks on quantized models can dramatically increase the damage \cite{rakin2019bit, yao2020deephammer, hong2019terminal}. 
However, it is \textit{not} always true that bit significance correlates to fault sensitivity. In PrioriFI \cite{weng2026priorifi}, it was shown that bit-level monotonicity (a more-significant bit is always more damaging) fails for a large fraction of pairwise comparisons on quantized edge NNs, once downstream nonlinearities mask the larger integer step. In studies like PrioriFI and FKeras \cite{weng2024fkeras} a thorough ranking of bits was determined via direct sensitivity measurements so that designers can efficiently target TMR or ECC on the most important subset of bits for selective hardware hardening. Other methods of determining selective hardening include BinFI, which locates SDC-causing bits by search \cite{chen2019binfi}; Aspis and related gradient heuristics rank weights \cite{schmedding2024aspis}. Beyond the passive hardening of designs, there are also active techniques as well. These methods include active checks on quantile markers for silent data corruption \cite{geissler2023low}, in-place zero-space ECC \cite{guan2019inplace}, functional ECC \cite{huang2020functional}, and weight-scaling \cite{fuengfusin2024harden, fuengfusin2024enhancing}.

Another direction in hardening neural networks is to teach them the environment they will be deployed in at training time, called Fault-Aware Training (FAT). Fault-injection frameworks like Li \textit{et al.}'s error-propagation study \cite{li2017understanding} and Ares \cite{reagen2018ares} have laid the foundation in software-simulated bit flips. These injection methods are then folded in during training \cite{zahid2020fat, gambardella2019efficient, vinck2025mitigating, malhotra2024memory}. The most related FAT strategy is Stutz \textit{et al.} \cite{stutz2021bit} where they trained quantized networks against random bit errors in stored weights and combined that with weight clipping. Cavagnero \textit{et al.}\ harden networks by fault-aware training and clipped activations at zero inference storage \cite{santos2024diehardnet}, and Chen \textit{et al.}\ restrict ranges after training \cite{chen2020ranger}. \ifsdsscolor Table~\ref{tab:heal_contribution} includes standalone unaligned surrogate FAT, with the alignment penalties and both runtime heals removed. Unaligned weight-bit-flip FAT and in-place zero-space ECC are still not arms.\else Neither unaligned weight-bit-flip FAT nor in-place zero-space ECC is an arm in our tables.\fi The goal is to regulate and harden the representation.

There have also been successful implementations of co-designing hardware with DNNs. ECON-T demonstrates a rad-tolerant NN ASIC \cite{diguglielmo2021reconfigurable} that runs a quantized NN while supporting full datapath TMR plus full-module TMR with autocorrect on the I$^{2}$C weight store on a real beam line \cite{braga2024econ}. That ASIC is the specialized-hardware reference for true TMR. The complementary case, and the target of this paper, is an unmodified commodity processor. 

\section{Problem Statement}
\label{sec:problem}
With all the major advances towards safe, reliable execution of DNNs against radiation faults, there remains an underserved sector. Particularly, how one might cheaply, on commodity commercial off-the-shelf (COTS) hardware, protect quantized neural networks against MBUs with a memory-light, architecture-independent method that preserves internal representation magnitudes critical for regression tasks.

\section{Methods}
The repair is arithmetic on the weights and activations already used at inference, so it runs as software on generic compute rather than in specialized radiation-hardened silicon. We outline that solution in three parts. First, we lay the foundations by discussing our fault injection simulation in Section~\ref{sec:mbu_sim} as groundwork for how we benchmark our models. Then we propose our new technique of \textit{Projection Healing} that actively corrects the weights and activations at runtime in Section~\ref{sec:topological_healing}. Alongside runtime protection, we also discuss FAT curriculum training that builds on top of established practice with an alignment penalty that routes signal to where the healing mechanism can recover it in Section~\ref{sec:fault_aware_training}. 

\subsection{MBU Simulation}
\label{sec:mbu_sim}
To simulate in software the effects of bit flips that would occur in hardware, we proceed by corrupting the binary representation of stored parameters. We assume a \emph{row-major fan-in} storage layout common for PyTorch \texttt{Linear.weight} tensors. Here a strike location is picked uniformly across the bit address and a contiguous burst of $m$ bits is flipped. Since the multiplicity $m$ is not fixed and depends on a variety of parameters (incident angle, energy, materials, etc.) we simplify the modeling by assuming it follows a truncated discrete power law of $P(m)\propto m^{-\alpha}$ on $\{1,\dots,m_{\max}\}$, with $\alpha=2.2$ and $m_{\max}=256$. We motivate the selection of the distribution's parameters by setting $33\%$ of strikes to be multi-bit ($> 1$ bit) which amounts to $\alpha=2.2$. We read the checkerboard entries in Table~II of \cite{ibe2010impact} as the share of upset events that are multi-cell, $\mathrm{MCU}/(\mathrm{MCU}+\mathrm{SEU})$, and match that share to $P(m\ge 2)$. At our operating point those odds are $P(m\ge 2)/P(m=1)\approx 0.49$, while $P(m\ge 2)=0.329$. The mean multiplicity is $\mathbb{E}[m]=2.65$ and the median multiplicity is $1$. Here these numbers are derived from the study's CORIMS simulator. Appendix~\ref{app:multiplicity} details how sensitive these results are to our choice in parameters.

At injection time, we corrupt stored INT8 weights and per-tensor INT8 biases. Per-channel weight scales (used to de quantize) remain protected float metadata and are not struck for our experiments. Biases are one-dimensional and are not passed through the row projection. The activation heal is their downstream correction, together with weight faults the weight gate misses, including the structurally inert $\mathrm{fc}_1$ weight heal. Transient activations are never injected. When benchmarking, the corrupted INT8 data is dequantized into floating point right before inference.

\subsection{Latent Space Redundancy: Projection Healing}
\label{sec:topological_healing}

\begin{figure}[t]
\centering
\includegraphics[width=\textwidth]{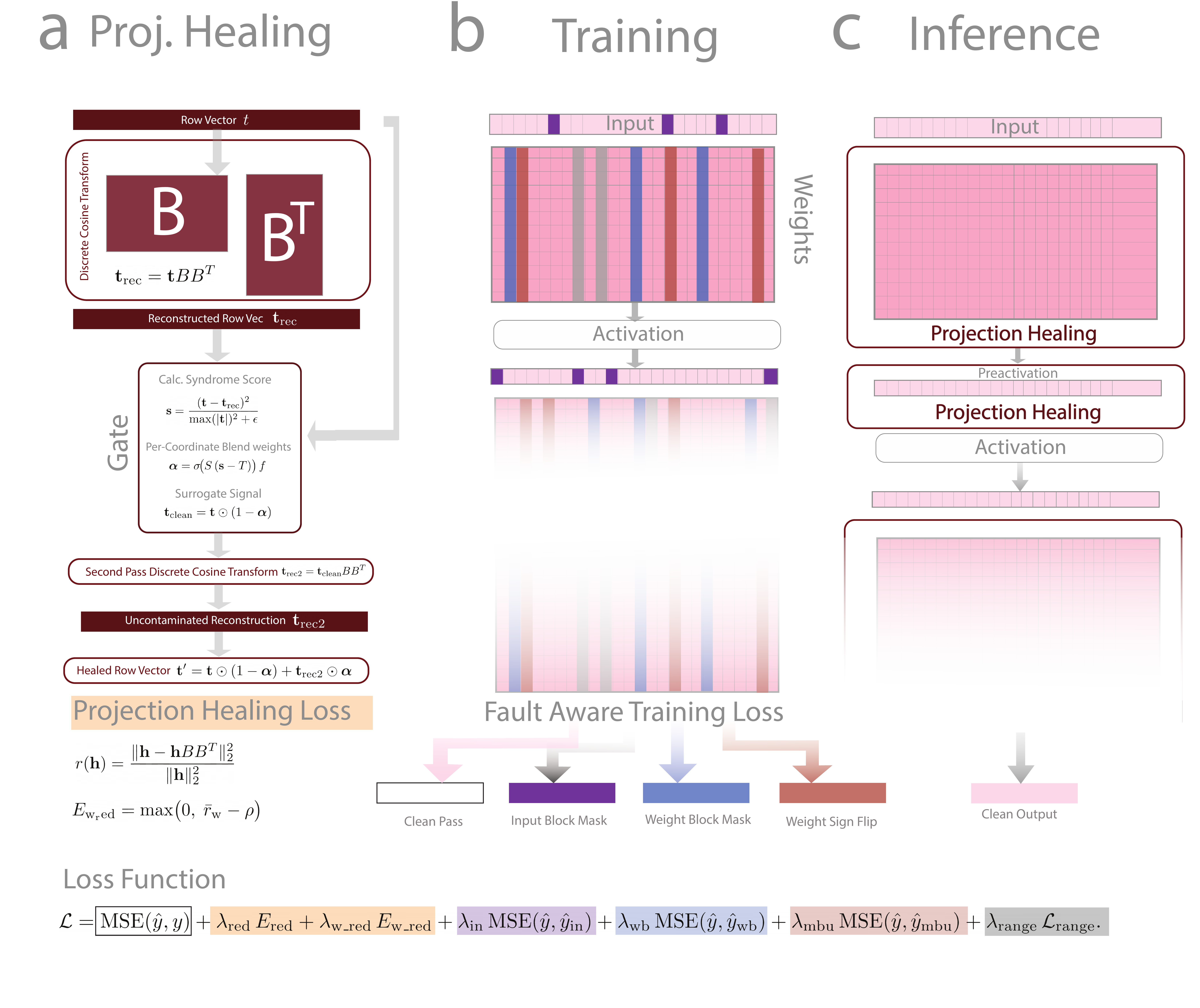}
\caption{System architecture. (a)~Two-pass projection healing on a row. The first DCT-II reconstruction forms the syndrome, the gate sets the blend weights, and a second projection of the surviving coordinates supplies the repaired entries. The weight-alignment term shown is $E_{\mathrm{w\_red}}=\max(0,\bar{r}_{\mathrm{w}}-\rho)$. (b)~Training uses four forwards: a clean pass, a contiguous input-block mask, a contiguous weight-block mask, and a contiguous weight sign flip. The range penalty is analytic on the weight rows, and the loss keeps the projection-healing terms separate from the fault-aware terms. (c)~A hidden layer heals its weight rows, heals the pre-activation, and then applies the nonlinearity. The drawing continues through the next weight matrix to the clean output.}
\label{fig:architecture}
\end{figure}

To achieve \textit{active} corrections live during runtime, we implement the idea of \textit{Projection Healing}. Projection healing's primary goal is to detect corrupted vectors and rebuild affected coordinates under the assumption they should live in a low-rank subspace shown in Figure~\ref{fig:architecture}a,c. This mechanism wraps around the linear layer backbone of a standard MLP and it requires no additional parameters or restrictions on layer width. We achieve this by aligning, at training, clean representations into the low-rank ``Green space'' so that injected fault noise is left occupying the orthogonal complement denoted as a ``Red space''. 

The choice in this low-rank subspace needs to be made ahead of time and also needs to be chosen so it can be reliably computed on the fly at runtime without dependence on stored parameters. To do this we chose the type-II Discrete Cosine Transform (DCT-II). DCT-II satisfies the constraint of providing a reliable runtime computable/checksummable basis while the type-II was chosen as it is the most common published type. We know we can do this because neuron ordering in an MLP is arbitrary and any fixed rank-$k$ orthogonal basis can align with the data should training forcefully route the model in such a way (Section~\ref{sec:mad_comparison}).  This is where the FAT alignment penalty comes into play (Section~\ref{sec:fault_aware_training}) where we force $95\%$ of clean activation signal into the low-rank subspace. For each layer we set a fixed Green rank $K=32$, forming $B\in\mathbb{R}^{N\times k}$ with $k=\max(1,\min(K,N-1))$. Appendix~\ref{app:k_sweep} sweeps this dimension against the clean-accuracy tax.

\subsubsection{Robust Projection}
To actually carry out the projection mechanism, we understand that any single injection only weakly changes the direction of the whole vector. Thus a standard projection would retain some fraction of the perturbation post-projection reconstruction which can continue to poison downstream inference. Thus we implement a robust single-pass projection for each row in the weight matrix and output hidden vector $\mathbf{t}\in\mathbb{R}^{1\times N}$. The robust projection first forms a provisional reconstruction $\mathbf{t}_{\mathrm{rec}}=\mathbf{t}BB^{T}$ and a per-coordinate syndrome normalized by the peak amplitude of the row:
\begin{equation}
\mathbf{s} = \frac{(\mathbf{t}-\mathbf{t}_{\mathrm{rec}})^{2}}{\max(|\mathbf{t}|)^{2}+\epsilon}.
\end{equation}
In our implementation we use $\epsilon=10^{-12}$ in this denominator, $10^{-6}$ in the concentration test, $10^{-10}$ in the red-energy ratio, and $10^{-8}$ as the range-penalty floor.
We then define a row-level master switch which is a function $f$ and a coordinate-level correction $\alpha_i$. Firstly,  a row is flagged only when its peak score is both large and concentrated:
\begin{equation}
f = \mathbb{I}\bigl(\max(\mathbf{s}) > \tau_{\mathrm{flag}} \;\land\; \max(\mathbf{s}) > 2\,\mathrm{mean}(\mathbf{s})+\epsilon\bigr),
\end{equation}
with $\tau_{\mathrm{flag}}=0.18$. However, we know a priori that certain coordinates of this vector need to be corrected; we can surgically make this correction by applying a per-coordinate gate $\alpha_i$ defined by a scaled and shifted sigmoid $\sigma$. 
\begin{equation}
\alpha_i = \sigma\!\bigl(S\,(\mathbf{s_i}-T)\bigr)\,f,
\end{equation}
For our experiments we use $(\tau_{\mathrm{flag}},T,S)=(0.18,0.25,40)$ for activation healing, and weight-row healing uses $(0.18,0.35,40)$. Appendix~\ref{app:gate_sweep} sweeps the weight-gate center parameters and marks $T=0.35$ as that operating point. We chose $S=40$ where the logistic is a near-hard switch on both paths. The DCT-II basis is a closed form of $(N,k)$, is checksummable, and can be regenerated rather than stored as persistent model state. Flagged coordinates are down-weighted to form a surrogate $\mathbf{t}_{\mathrm{clean}}=\mathbf{t}\odot(1-\boldsymbol{\alpha})$, from which an uncontaminated reconstruction $\mathbf{t}_{\mathrm{rec2}}=\mathbf{t}_{\mathrm{clean}}BB^{T}$ is derived, yielding the healed row:
\begin{equation}
\mathbf{t}' = \overbrace{\mathbf{t}\odot(1-\boldsymbol{\alpha})}^{\text{original healthy coord.}} + \underbrace{\mathbf{t}_{\mathrm{rec2}}\odot\boldsymbol{\alpha}}_{\text{corrected corrupt coord. }}
\end{equation}

The intuition behind the need for a double reconstruction is that if we used a single projection it would both \textit{see} the spike and \textit{keep} a piece of it. So we need the second pass there so the repair is constructed from the coordinates that were \textit{not} corrupted. If we projected once, it would have mixed together corrupted signal instead of surgically removing it. However, during training, projection healing is \textit{not} active and instead there is a simple alignment regularization applied instead.

\subsubsection{Weight Healing}
Now that we have a mechanism for healing, at inference we apply it to every weight row before the linear map, including the readout. Healed values are not written back to the INT8 SRAM and instead they live in the float working copy that dequantizing inference already materializes. The heal therefore sees whatever INT8 damage has accumulated since the last golden-copy reload, and it repairs only the coordinates the gate can flag, not the whole image. In the benchmark that snapshot is injected in one go. Because nothing is written back, that one-shot dump is the same stored state as strikes arriving over time on SRAM. We do not model faults in the float working copy, nor a stream of strikes interleaved with inference. For sequential inference, rather than retaining that working copy, the healed INT8 image can be written back on a tunable cadence. Section~\ref{sec:compute_runtime} and Appendix~\ref{app:overhead} give the cost of one writeback every $Q=256$ iterations.

\subsection{Fault Aware Training}
\label{sec:fault_aware_training}
To turn the projection healing live we need to apply FAT to our model. Firstly we build on top of established FAT practice \cite{zahid2020fat, vinck2025mitigating, santos2024diehardnet}. Since MBUs cascade across adjacent bits, each corrupted forward draws one contiguous block, not independent Bernoulli dropout and not a two-dimensional sub-block. The input mask zeros a contiguous $25\%$ of features. Separately, every linear layer from $\mathrm{fc}_1$ through $\mathrm{fc}_{\mathrm{out}}$ receives one contiguous fan-in block: zeros on $25\%$ of columns, or, on a different forward, sign inversion on $15\%$ of columns. The weight-block and sign-flip passes are not applied together. This ensures we do not overly rely on any particular input feature for estimation. We note that this is mismatched to how our evaluator injects faults. In fact, these simulate spatial MBUs but are not true INT8 bit flips; they do not follow the power-law multiplicity (Appendix~\ref{app:multiplicity}). This was intentionally done so that training never observes the fault distribution it is graded against and cannot overfit to a particular process. We also verified in Appendix~\ref{app:bitflip_fat} whether this mismatch leaves performance on the table and we observe a small $3.3\%$ difference in mean excess AUC at the nearest available curriculum severity. In direct comparison to DieHardNet \cite{santos2024diehardnet} which corrupts intermediate feature maps for noise tolerance, we corrupt structural weights to simulate SRAM and DRAM flips directly. At training time to ensure stability we regularize the network by the squared deviation between the clean prediction $\hat{y}$ and predictions under input-block masking ($\hat{y}_{\mathrm{in}}$), weight-block dropout ($\hat{y}_{\mathrm{wb}}$), and sign flips ($\hat{y}_{\mathrm{mbu}}$).

Beyond the standard fault injection we also have the important \textbf{alignment penalty}.  First, we probe individually the \textit{pre-activations} and quantify on average how much of the vectors live outside of the ``green'' subspace $\bar r_{act}$.  By first computing the normalized orthogonal syndrome for a $\mathbf{v}\in\mathbb{R}^{1\times N}$ against the assumed DCT-II basis $B\in\mathbb{R}^{N\times k}$,
\begin{equation}
r(\mathbf{v}) = \frac{\|\mathbf{v}-\mathbf{v}BB^{T}\|_{2}^{2}}{\|\mathbf{v}\|_{2}^{2}+\epsilon}.
\end{equation}
A minibatch of clean pre-activations $H^{(\ell)}\in\mathbb{R}^{n_{\mathrm{mb}}\times N_{\ell}}$ is scored with the fan-out basis of length $N_{\ell}$,
\begin{equation}
\bar{r}_{\mathrm{act}}^{(\ell)} = \frac{1}{n_{\mathrm{mb}}}\sum_{i=1}^{n_{\mathrm{mb}}} r\bigl(H^{(\ell)}_{i,:}\bigr).
\end{equation}
Next, we define a similar metric for the \textit{weights}. Depth weights $w=(0.5,0.75,1.0)$ reduce early-layer pressure. The activation hinge in Eq.~(\ref{eq:ered}) sums over the three hidden layers $\ell\in\{\mathrm{fc}_1,\mathrm{fc}_2,\mathrm{fc}_3\}$ and excludes the scalar readout. Width scales are $s_{\ell}=\sqrt{N_{\min}/N_{\ell}}$ with $N_{\min}=\min_{\ell} N_{\ell}$ over those three hidden widths. Weight matrices $W^{(\ell)}\in\mathbb{R}^{N_{\mathrm{out}}\times N_{\mathrm{in}}}$ are scored row-wise with the fan-in basis of length $N_{\mathrm{in}}$, then averaged over output neurons and over $\mathcal{L}=\{\mathrm{fc}_2,\mathrm{fc}_3,\mathrm{fc}_{\mathrm{out}}\}$:
\begin{equation}
\bar{r}_{\mathrm{w}} = \frac{1}{|\mathcal{L}|}\sum_{\ell\in\mathcal{L}}\frac{1}{N_{\mathrm{out}}^{(\ell)}}\sum_{i=1}^{N_{\mathrm{out}}^{(\ell)}} r\bigl(W^{(\ell)}_{i,:}\bigr).
\end{equation}
With these quantified metrics $\bar{r}_{\mathrm{act}}$ and $\bar{r}_{\mathrm{w}}$ we use hinges above a dead zone $\rho=0.05$ to hold residual Red content at a measured operating point without degrading clean accuracy. The activation hinge loss is applied per layer before the weighted mean; the weight hinge is applied after averaging across layers giving us penalty metrics $E_{red}$ and $E_{w\_red}$. We note that layer $\mathrm{fc}_1$ is excluded from the weight-alignment sum due to small fan-in. 
\begin{equation}
E_{\mathrm{red}} = \frac{\sum_{\ell} w_{\ell}s_{\ell}\,\max\!\big(0,\;\bar{r}_{\mathrm{act}}^{(\ell)}-\rho\big)}{\sum_{\ell} w_{\ell}s_{\ell}},
\label{eq:ered}
\end{equation}
\begin{equation}
E_{\mathrm{w\_red}} = \max\!\big(0,\;\bar{r}_{\mathrm{w}}-\rho\big).
\end{equation}
We also constrain the weight ranges in the network via a \textbf{range penalty}. We want to ensure that the weights do not vary drastically, which could make the healing mechanism falsely identify corrupted bits. A range penalty is applied on every Linear layer, including $\mathrm{fc}_1$, by penalizing the ratio of peak to mean absolute weight, with the readout weighted three times more heavily than hidden layers:
\begin{equation}
\begin{aligned}
\mathcal{L}_{\mathrm{range}}
&=
\frac{1}{\sum_{\ell}\gamma_{\ell}}
\sum_{\ell}\gamma_{\ell}\,
\frac{1}{N_{\mathrm{out}}^{(\ell)}}
\sum_{i=1}^{N_{\mathrm{out}}^{(\ell)}}
\frac{\|W^{(\ell)}_{i,:}\|_{\infty}}{\frac{1}{N_{\mathrm{in}}^{(\ell)}}\|W^{(\ell)}_{i,:}\|_{1}+\epsilon},
\\
\gamma_{\ell}
&=
\begin{cases}
3 & \ell=\mathrm{fc}_{\mathrm{out}},\\
1 & \text{otherwise.}
\end{cases}
\end{aligned}
\end{equation}
Together, the full objective separates the alignment terms from the FAT terms:
\begin{align}
\mathcal{L}
&=
\mathrm{MSE}(\hat{y}, y)
\nonumber\\
&\quad
+
\underbrace{
\lambda_{\mathrm{red}}\, E_{\mathrm{red}}
+
\lambda_{\mathrm{w\_red}}\, E_{\mathrm{w\_red}}
}_{\text{Projection Healing}}
\nonumber\\
&\quad
+
\underbrace{
\begin{aligned}[t]
&\lambda_{\mathrm{in}}\,
\mathrm{MSE}(\hat{y}, \hat{y}_{\mathrm{in}})
+
\lambda_{\mathrm{wb}}\,
\mathrm{MSE}(\hat{y}, \hat{y}_{\mathrm{wb}})
\\[-0.15em]
&{}+
\lambda_{\mathrm{mbu}}\,
\mathrm{MSE}(\hat{y}, \hat{y}_{\mathrm{mbu}})
+
\lambda_{\mathrm{range}}\, \mathcal{L}_{\mathrm{range}}
\end{aligned}
}_{\text{Fault-Aware Training}}.
\end{align}
FAT coefficients scale dynamically via an exponential moving average of task MSE, and $\lambda_{\mathrm{red}}$ is annealed over the first $40$ epochs. The six base $\lambda$ multipliers were tuned once on a development split to balance clean accuracy against alignment, and kept fixed across all datasets.
For our work we chose to use $\lambda_{\mathrm{red}}^{\mathrm{end}}=8$ as the plateau values for $\lambda_{\text{red}}$. Here the red coefficient ramps linearly, $\lambda_{\mathrm{red}}(e)=8\min(e/40,1)$, and is not decayed afterward.  We set other parameters $\lambda_{\mathrm{w\_red}}=0.5$, $\lambda_{\mathrm{in}}=0.05$, $\lambda_{\mathrm{wb}}=0.05$, $\lambda_{\mathrm{mbu}}=0.015$, and $\lambda_{\mathrm{range}}=0.01$. Both the alignment group and the FAT group are multiplied by an exponential moving average (EMA) scale factor $s=\max(\mathrm{EMA}/0.25,\,10^{-3})$, where $\mathrm{EMA}\leftarrow 0.99\,\mathrm{EMA}+0.01\,\mathrm{MSE}$ and the reference MSE is $0.25$. Training is AdamW at learning rate $10^{-3}$, weight decay $10^{-5}$, minibatch size $256$, dropout $0.1$, LeakyReLU negative slope $0.01$, and gradient clip $1$. Optimization runs for at most $500$ epochs with ReduceLROnPlateau (factor $0.5$, patience $5$, floor $10^{-6}$) and keeps the best validation checkpoint.
\subsection{Benchmarking Protocol}
\label{sec:benchmarking}
For our benchmarking test, we use six regression-specific datasets. The first two are generic tabular benchmarks and the remaining four are domain problems. The generic tabular datasets are \textbf{California Housing} \cite{pace1997sparse}, which predicts house values, and \textbf{Friedman \#1} \cite{friedman1991multivariate} ($N{=}10{,}000$), which provides a nonlinear synthetic target. The other four datasets are \textbf{SDSS photometric redshift}, which predicts continuous $z$ from \ifsdsscolor the reference magnitude $r$ and the colors $u{-}g$, $g{-}r$, $r{-}i$, and $i{-}z$ \cite{beck2016photometric, collister2004annz}\else DR12 passband magnitudes \cite{beck2016photometric, collister2004annz}\fi, targeting spaceborne observability. \textbf{NASA C-MAPSS FD001} \cite{saxena2008damage, xiong2023controlled} predicts Remaining Useful Life from turbofan sensors. \textbf{NRC Critical Heat Flux} \cite{lecorre2024chf} predicts boiling-crisis thresholds from the public US NRC critical heat flux tube database (NUREG/KM-0011). \textbf{CMS Jet $\boldsymbol{p_T}$} regresses generator-level jet transverse momentum from the EnergyFlow CMS2011AJets MOD HDF5 simulation sample (companion to the CMS 2011A Open Data jet record) \cite{cms2011a_jet_opendata, zenodo_cms2011a_sim470, komiske2020exploring}, an off-detector trigger-side HEP task \cite{duarte2018fast}. Per-dataset architectures and INT8 footprints are detailed in Appendix~\ref{app:architectures}.

Features and targets are standardized for training. Predictions are inverse-transformed to native units for reporting RMSE. Post-training, weights are quantized to symmetric per-channel INT8; activations remain floating point.

\subsubsection{Evaluation Metrics}
\label{sec:evaluation_metrics}
We define the MBU \emph{intensity} $I$ as the fraction of corrupted weight bits since the last scrub \begin{equation}
I = \frac{N_{\mathrm{clusters}} \times \mathbb{E}[m]}{N_{\mathrm{params}} \times 8},
\label{eq:intensity}
\end{equation}
where $m$ follows the truncated power law of Eq.~(\ref{eq:multiplicity}) at $\alpha{=}2.2$, $m_{\max}{=}256$, and $\mathbb{E}[m]\approx 2.65$.

We report three metrics.  Firstly, \textbf{RMSE degradation} evaluates held-out RMSE against a logarithmically swept $I$. Second, \textbf{Operational and breakdown thresholds} are the intensities at which median RMSE rises $+5\%$, $+10\%$, or $+50\%$ above each model's own clean baseline. Here $I_{\mathrm{break}}$ means this $+50\%$ own-baseline threshold. And third, \textbf{Relative excess AUC} integrates $\max(y-y^\star,0)$ against $\log_{10}I$ for $I\in[10^{-4},1]$, scaled by the unprotected curve's integral, where $y^\star$ is the best clean RMSE across models. A value below $1$ indicates less excess error. Note that some results  report the unnormalized native-unit excess $\int\max(y-y^\star,0)\,d\log_{10}I$, or the absolute integral $\int y\,d\log_{10}I$ as opposed to the relative in order to characterize absolute performance. Those are identified in each caption. Thresholds use each model's own clean RMSE; excess AUC uses the shared floor $y^\star$. The two conventions diverge when the alignment tax already exceeds the operational tolerance.

Each intensity data point represents a median of $400$ independent MBU realizations. The tables report mean$\pm$standard deviation across $8$ seeds of each per-seed median of $400$ individual realizations.

\section{Results}
To gauge the performance of our design we investigate the benchmarking results first in Section \ref{sec:results_sweeps} and then isolate the mechanisms in Sections \ref{sec:heal_attribution}, \ref{sec:mad_comparison}, \ref{sec:single_comp_rem}, and \ref{sec:leakage}. Lastly, we investigate bit-level sensitivity in Section \ref{sec:silent_band_summary}.

\subsection{Benchmarking Results}
\label{sec:results_sweeps}

\begin{figure}[!t]
\centering
\ifsdsscolor
\includegraphics[width=\textwidth,height=0.48\textheight,keepaspectratio]{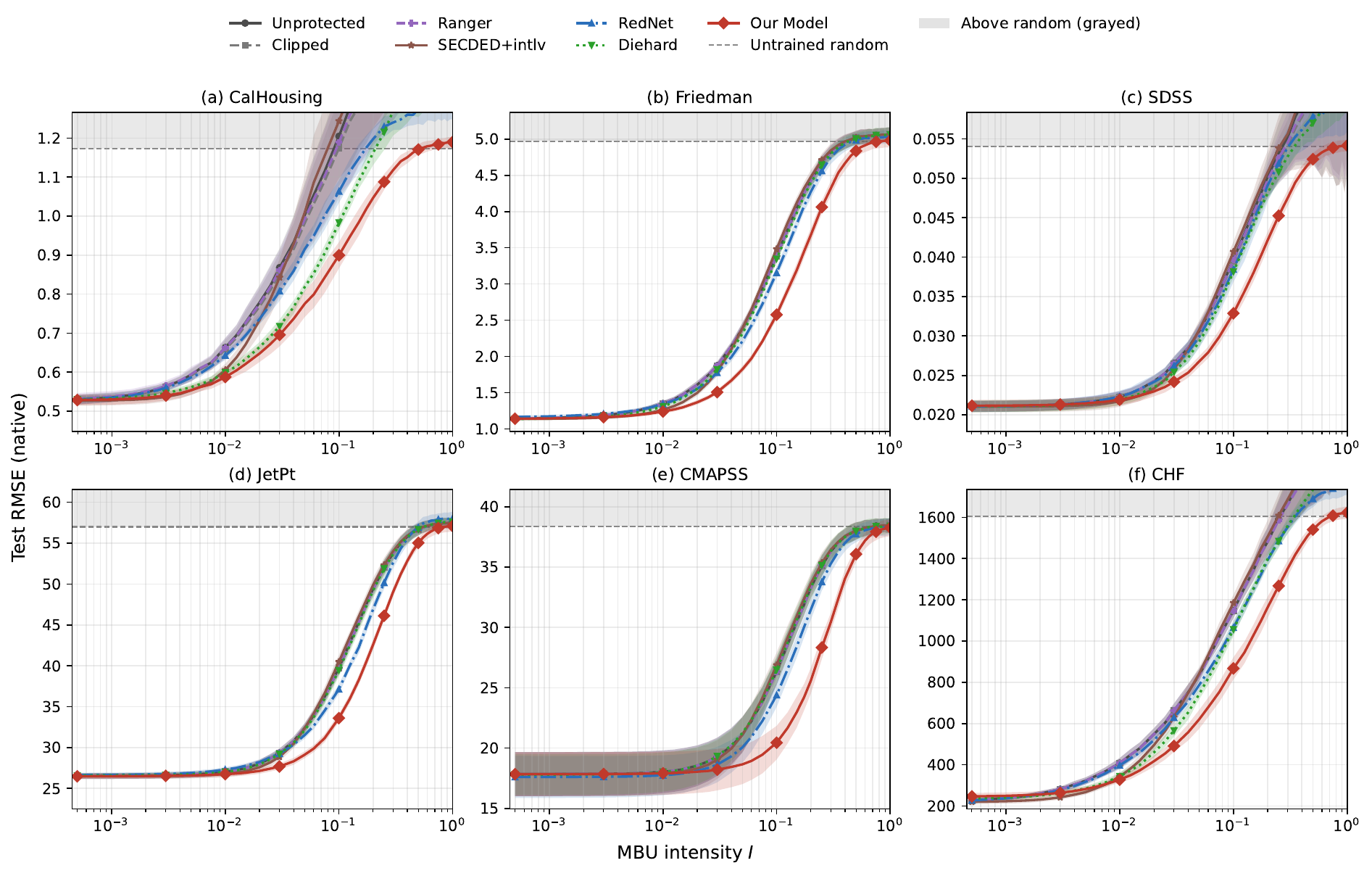}
\else
\includegraphics[width=\textwidth,height=0.48\textheight,keepaspectratio]{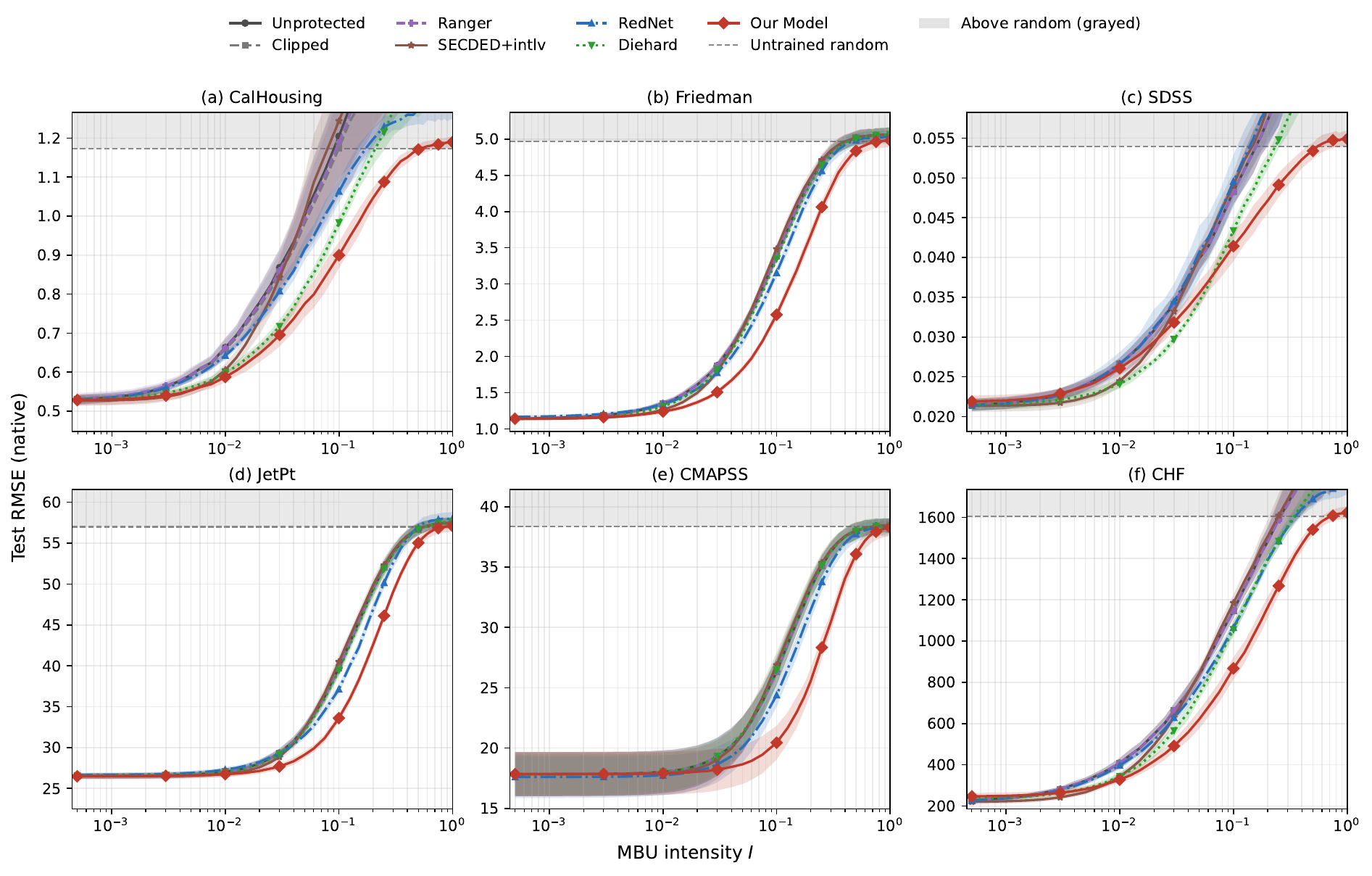}
\fi
\ifsdsscolor
\caption{MBU sweeps on six datasets ((a)--(f)). Log x-axis: accumulated weight-bit fraction $I$ over $[10^{-4},1]$. Linear y-axis: held-out test RMSE in native target units (mean across $8$ seeds; shaded bands are across-seed std). Curves: unprotected, Clipped, Ranger~\cite{chen2020ranger}, SECDED+interleaving, RedNet, Diehard, and Our Model (RedNet/Diehard are proxy implementations; Appendix~\ref{app:baselines}). The dashed line is the untrained-random floor and the gray band is the uninformative region above it. Ranger matches Clipped by construction under INT8 (Appendix~\ref{app:baselines}). SECDED dominates at low $I$ then saturates; Our Model stays below the software defenses longest. The SDSS panel uses the color features of Section~\ref{sec:benchmarking}.}
\else
\caption{MBU sweeps on six datasets ((a)--(f)). Log x-axis: accumulated weight-bit fraction $I$ over $[10^{-4},1]$. Linear y-axis: held-out test RMSE in native target units (mean across $8$ seeds; shaded bands are across-seed std). Curves: unprotected, Clipped, Ranger~\cite{chen2020ranger}, SECDED+interleaving, RedNet, Diehard, and Our Model (RedNet/Diehard are proxy implementations; Appendix~\ref{app:baselines}). The dashed line is the untrained-random floor and the gray band is the uninformative region above it. Ranger matches Clipped by construction under INT8 (Appendix~\ref{app:baselines}). SECDED dominates at low $I$ then saturates; Our Model stays below the software defenses longest.}
\fi
\label{fig:mbu_sweeps}
\end{figure}

We first investigate trends in model performance as a function of intensity. In Figure~\ref{fig:mbu_sweeps} we see across the 6 regression benchmarks that all models trend towards higher error rates monotonically. The unprotected quantized INT8 network consistently degrades the most rapidly, while comparative benchmark models like RedNet and Diehard (implementations of the published methods; see Appendix~\ref{app:baselines}) provide intermediate protection. Likewise with classically protected ECC weights (SECDED+Interleaved) but all models collapse once MBUs become dominant at higher $I$. Nearly all models later become indistinguishable from an untrained network at $I \sim 0.3$--$0.4$. Thus characterizing performance is really about the \textit{approach} towards that collapse.
Beyond visual comparisons, we can quantify the improvement by computing the relative excess area under the curve (AUC). Table~\ref{tab:rel_auc_combined} shows two stacked panels (a)~the early operational window ($I \in [10^{-4}, 10^{-2}]$) and (b)~the wide window ($I \in [10^{-4}, 1]$).  We see that in the early operational stage ($I \in [10^{-4}, 10^{-2}]$) the effects are nearly flat. This is due to the effects being dominated by SEU-like disruptions and is thus a prime setting for classical algorithms like SECDED with interleaving to shine, producing a median excess AUC of $0.49$ vs. our proposed model's $0.76$. The operational-window mean for our method is $1.11$, worse than the unprotected baseline of $1.00$, because SDSS and CMAPSS dominate the average. On CMAPSS the unprotected excess over this window is near zero, so a small clean gap above the shared floor $y^\star$ inflates the ratio. The median, $0.76$, is the summary we use for that window. However we see that this picture changes once we consider the complete window ranging from $I \in [10^{-4}, 1]$. Here when MBU becomes a dominant source of error we see that our model still maintains a wide-window mean \ifsdsscolor\sdssnew{$0.71$ (median $0.72$)}\else$0.70$ (median $0.71$)\fi\ vs.\ SECDED $0.99$  where single-bit correction is insufficient. We also see that the other models, DieHard and RedNet, are also effectively inert at ${\sim}0.9$ in the wide-window regime but still yield meaningful gains over a standard ECC or unprotected network as they incorporate FAT and specialized layer ordering to combat these faults. 

\input{tables/rel_auc_combined}

We can also measure the breakdown and degradation thresholds in Table~\ref{tab:breakdown_full}  
\input{tables/breakdown_full} which reports the intensities at which median RMSE rises $+5\%$, $+10\%$, and $+50\%$ above each model's own clean baseline. We see that at $+5\%$ we extend the median interval by ${\sim}2\times$ over the unprotected baseline so the healing mechanism does suppress early dominant errors similar to that of the classical SECDED error-correcting method. However we see classical performance drop as the model experiences more faults as SECDED eventually loses its lead at the $10$--$50\%$ degradation stages while our method maintains the $2\times$ lead over unprotected. 

Despite these gains in robustness, routing capacity into the low-rank subspace incurs an ``alignment tax'' on fault-free held-out RMSE, $(\mathrm{Ours}-\mathrm{Unprot})/\mathrm{Unprot}$. Across the six datasets the median tax is $+0.3\%$ and the mean is ${\sim}2\%$, with CHF the only material outlier ($+11.8\pm5.8\%$); the other five sit within a few percent and are consistent with zero within seed scatter. CHF therefore begins at $I=0$ already beyond the unprotected model's $+5\%$ and $+10\%$ RMSE budgets. Its own-baseline $+5\%$ intensity is a drift from that degraded clean floor. Against the shared floor $y^\star$ the stored CHF means give $+12.4\%$, and the shared-floor $+5\%$ threshold is $I=0$ on all eight seeds. We do not convert the CHF own-baseline threshold into a calendar lifetime at the unprotected clean RMSE. Relative to the $10$--$50\%$ RMSE degradations that define the operational windows, this trade-off is negligible except on CHF.

\subsection{Ablation Results}
\label{sec:ablation_results}
Given the improvements over other baseline techniques, we seek to pin down the weight-bearing mechanism providing the performance improvements. We first check the projection healing mechanism over FAT in Section \ref{sec:heal_attribution}, then the basis and MAD comparison (Section~\ref{sec:mad_comparison}), then single-component removals (Section~\ref{sec:single_comp_rem}). Per-bit gate coverage is Section~\ref{sec:silent_band_summary}.

\subsubsection{Contribution of the Runtime Heal Beyond Fault-Aware Training}
\label{sec:heal_attribution}
Since it has been established in the literature that FAT can substantially improve model robustness, we isolate the two possibly confounded mechanisms from each other by ablating the two. We isolate the projection heal by holding a fully FAT-trained checkpoint fixed and toggling projection healing to be off. Any performance difference is then due to the runtime mechanism. \ifsdsscolor The \sdssnew{$23.4\%$} figure is the mean of $(\mathrm{No\ heal}-\mathrm{Full})/\mathrm{No\ heal}$ on the native-unit excess integral in Table~\ref{tab:heal_contribution}, from \sdssnew{$22.3\%$} on SDSS to $33.7\%$ on CMAPSS. Against standalone unaligned FAT the same full stack averages \sdssnew{$13.7\%$}, a tie on SDSS (\sdssnew{$-0.2\%$}) and \sdssnew{$25.4\%$} on CMAPSS. \else The $22.2\%$ figure is the mean of $(\mathrm{No\ heal}-\mathrm{Full})/\mathrm{No\ heal}$ on the native-unit excess integral in Table~\ref{tab:heal_contribution}, from $14.6\%$ on SDSS to $33.7\%$ on CMAPSS. \fi Clean $\Delta$ is the fault-free RMSE change from toggling the runtime heal on that same frozen checkpoint. We also see that removing the activation heal still hurts under weight-only injection because unflagged weight faults propagate into pre-activations where the activation gate can repair them.

\input{tables/heal_contribution}

\subsubsection{MAD Despike and Basis Swap}
\label{sec:mad_comparison}
To verify that our surgical healing mechanism with projections cannot be trivialized via simpler statistical methods, we implement a \textbf{MAD despiker.} By holding the FAT curriculum and alignment fixed, we replace the masked projection with per-row MAD median fill. Here the fixed training objective includes the $\mathcal{L}_{\mathrm{range}}$ penalty, which penalizes the ratio of row peak absolute weight to row mean absolute weight, aligning it to the MAD despiker mechanism by suppressing extreme clean feature values, protecting it from spuriously amputating legitimate heavy tails. We see that \textbf{MAD despiker} cannot replicate our performance in Table \ref{tab:arch_retrain} as, under the same absolute re-integration used for the basis swap, MAD is strictly worse on all six ($1.07\times$--$1.87\times$).

\textbf{Arbitrary fixed orthogonal basis.} We also verify whether our choice in the DCT-II basis is privileged and potentially ``overfitted'' to this problem by checking if we chose a random arbitrary orthogonal basis if we could replicate similar performance. Table~\ref{tab:arch_retrain} shows near-parity on four datasets. CHF is the exception: the arbitrary basis is $1.78\times$ worse, in the same direction as its $+11.8\%$ alignment tax. Our choice in DCT-II basis remains rooted not in performance gain but in the fact that it is an analytic closed form, that can be reproduced with bit-exactness and it can be audited against a published definition and checksummed in flight. The pseudorandom basis chosen for the experiment does not have such similar properties as either the basis needs to be stored (exposed to faults) or the basis needs to be orthogonalized on the fly where bit-exact regeneration is not guaranteed and can be slow compared to closed-form methods $O(Nk)$ closed-form cosines against Gaussian sampling followed by an $O(Nk^{2})$ orthogonalization.

\input{tables/arch_retrain}

\subsubsection{Single-Component Removals}
\label{sec:single_comp_rem}
Lastly, we investigate dropping single mechanisms and retraining models from scratch to see how they individually affect performance. In Table~\ref{tab:ablation_fractional_auc} we see that removing weight alignment is the single most damaging variation: the heal needs that aligned low-rank subspace. While removing the range penalty is slightly better on the mean. \ifsdsscolor We see that precision is better on Friedman, SDSS, JetPt, and CMAPSS, tied on CHF, and worse on CalHousing. The other removals degrade the mean. Here the no-weight-align arm has a mean \sdssnew{$1.83$}, which is the evidence that the runtime heal fails without the aligned subspace.\else We see that precision is better on Friedman, JetPt, and CMAPSS, tied on CHF, and worse on CalHousing and SDSS. The other removals degrade the mean. Here the no-weight-align arm has a mean $1.73$, which is the evidence that the runtime heal fails without the aligned subspace.\fi

\begin{table}[!t]
\centering
\caption{Normalized relative excess AUC, same definition as Table~\ref{tab:rel_auc_combined}, on knots from $5\times 10^{-4}$ to $1$. Lower is better. Each column is a from-scratch retrain except the unprotected reference.}
\label{tab:ablation_fractional_auc}
\setlength{\tabcolsep}{2.2pt}
\begin{tabular}{lccccccc}
\hline
Configuration & CalH. & Fried. & SDSS & JetPt & CMAPSS & CHF & Mean \\
\hline
Unprotected & 1.00 & 1.00 & 1.00 & 1.00 & 1.00 & 1.00 & 1.00 \\
Full (Ours) & $\mathbf{0.58\pm0.10}$ & $\mathbf{0.79\pm0.02}$ & \ifsdsscolor\sdssnew{$\mathbf{0.71\pm0.10}$}\else$\mathbf{0.63\pm0.09}$\fi & $\mathbf{0.78\pm0.02}$ & $\mathbf{0.69\pm0.04}$ & $\mathbf{0.73\pm0.07}$ & \ifsdsscolor\sdssnew{$\mathbf{0.71\pm0.06}$}\else$\mathbf{0.70\pm0.06}$\fi \\
No concentration flag & $0.58\pm0.09$ & $0.80\pm0.03$ & \ifsdsscolor\sdssnew{$1.08\pm0.18$}\else$0.75\pm0.11$\fi & $0.78\pm0.01$ & $0.69\pm0.04$ & $0.80\pm0.08$ & \ifsdsscolor\sdssnew{$0.79\pm0.07$}\else$0.73\pm0.06$\fi \\
No act.\ heal & $0.64\pm0.09$ & $0.83\pm0.03$ & \ifsdsscolor\sdssnew{$0.77\pm0.11$}\else$0.69\pm0.09$\fi & $0.82\pm0.02$ & $0.70\pm0.04$ & $0.78\pm0.07$ & \ifsdsscolor\sdssnew{$0.76\pm0.06$}\else$0.74\pm0.06$\fi \\
No weight heal & $0.60\pm0.09$ & $0.89\pm0.02$ & \ifsdsscolor\sdssnew{$0.75\pm0.11$}\else$0.64\pm0.09$\fi & $0.86\pm0.02$ & $0.81\pm0.06$ & $0.79\pm0.07$ & \ifsdsscolor\sdssnew{$0.78\pm0.06$}\else$0.76\pm0.06$\fi \\
No weight align & $1.06\pm0.18$ & $2.21\pm0.13$ & \ifsdsscolor\sdssnew{$1.86\pm0.29$}\else$1.22\pm0.17$\fi & $2.54\pm0.26$ & $1.79\pm0.35$ & $1.54\pm0.12$ & \ifsdsscolor\sdssnew{$1.83\pm0.22$}\else$1.73\pm0.20$\fi \\
No act.\ red & $0.64\pm0.10$ & $1.28\pm0.10$ & \ifsdsscolor\sdssnew{$0.82\pm0.13$}\else$0.61\pm0.09$\fi & $1.07\pm0.07$ & $0.91\pm0.09$ & $0.95\pm0.07$ & \ifsdsscolor\sdssnew{$0.94\pm0.09$}\else$0.91\pm0.09$\fi \\
No MBU loss & $0.60\pm0.10$ & $0.80\pm0.02$ & \ifsdsscolor\sdssnew{$0.69\pm0.10$}\else$0.62\pm0.09$\fi & $0.78\pm0.01$ & $0.70\pm0.06$ & $0.74\pm0.07$ & \ifsdsscolor\sdssnew{$0.72\pm0.06$}\else$0.71\pm0.06$\fi \\
No range pen. & $0.65\pm0.10$ & $0.77\pm0.02$ & \ifsdsscolor\sdssnew{$0.68\pm0.10$}\else$0.67\pm0.10$\fi & $0.71\pm0.04$ & $0.64\pm0.06$ & $0.73\pm0.07$ & \ifsdsscolor\sdssnew{$0.70\pm0.06$}\else$0.69\pm0.06$\fi \\
\hline\end{tabular}
\end{table}

\subsubsection{Generalization to Unseen Fault Models}
\label{sec:leakage}
We replace the surrogate FAT curriculum with an exact evaluation fault injector, to check that the reported resilience is not an artifact of matching the training curriculum to the evaluation distribution (Appendix~\ref{app:bitflip_fat}). We see that across the six-dataset geometric mean excess AUC differs by only $3.3\%$. Note that this $3.3\%$ does not measure unaligned weight-bit-flip FAT. The no-heal column of Table~\ref{tab:heal_contribution} still contains the alignment penalties. The no-weight-align row of Table~\ref{tab:ablation_fractional_auc} removes subspace alignment but is still our surrogate curriculum with eval-time healing.

\subsection{Per-Bit-Position Damage and Detection}
\label{sec:silent_band_summary}
\begin{figure}[!t]
\centering
\includegraphics[width=\textwidth]{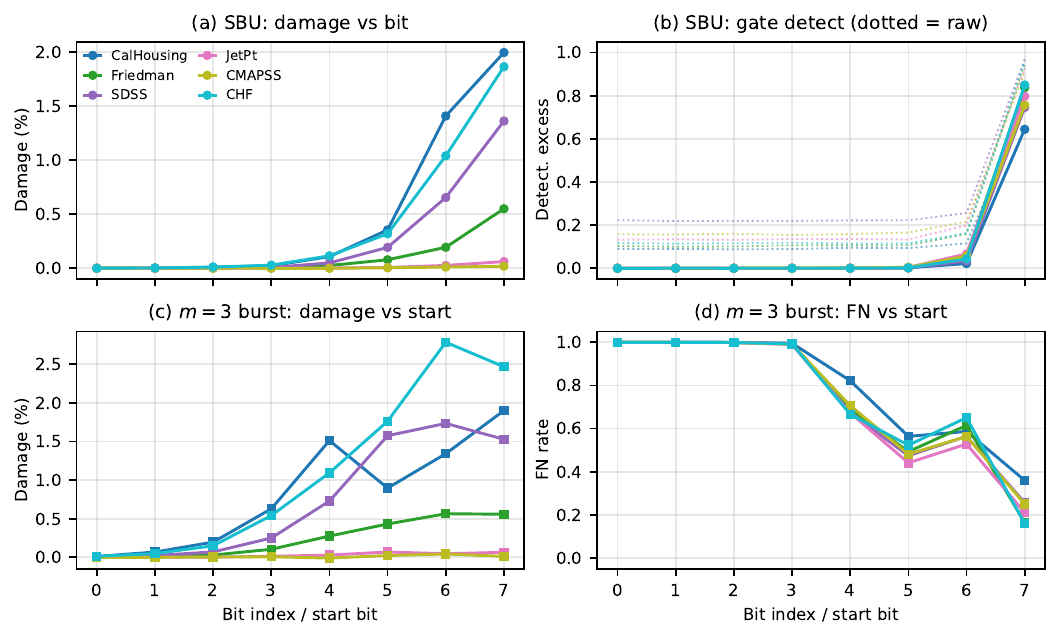}
\caption{Per-bit-position curves on the six published checkpoints. (a)~SBU relative RMSE damage vs.\ bit index. (b)~SBU row-level gate detection excess vs.\ bit index (dotted: raw fire rate). (c)--(d)~Short bursts ($m=3$) with forced start bit: relative damage and baseline-corrected false-negative rate vs.\ start bit. Low-order bits are nearly silent; the sign bit dominates both damage and detection.}
\label{fig:bitpos}
\end{figure}

Since half the battle of designing a healing mechanism is \textit{detecting} where and when to correct a parameter, one key metric is to uncover how sensitive the mechanism is to weight corruption at the bit level. In Figures~\ref{fig:bitpos}(a) and \ref{fig:bitpos}(c), bits $0$--$5$ contribute little SBU damage and bits $6$ and $7$ dominate. The parenthetical values $<0.01$ and $0.04$ are gate-detection excesses from Figure~\ref{fig:bitpos}(b) and Table~\ref{tab:gate_threshold_sweep}, not the damage axis of Figure~\ref{fig:bitpos}(a). Under a fixed burst width $m=3$,  the sign bit accounts for most of the plotted damage. Pooling the six-dataset mean relative-RMSE damages it gives a share of $12.8\%$ (bits $0$--$5$), $31.6\%$ (bit~$6$), and $55.5\%$ (bit~$7$). This makes sense as under INT8, bit~$7$ is the largest step in the word, so we calibrated the gate to the band where active reconstruction is both feasible and consequential in Figures~\ref{fig:bitpos}(b) and \ref{fig:bitpos}(d). The sharp change between bits~$6$ and~$7$ is attributed to the syndrome algebra of Section~\ref{sec:topological_healing}. A coordinate is corrected only when its row is flagged \emph{and} the coordinate itself clears the gate center, $f\land(s_i>T)$.

However, this would call into question the need for projection healing if only a few bits require real protection, which should be easily achieved via classical ECC. To address this we benchmark this specific case with a dedicated SECDED(b7) arm in Table~\ref{tab:rel_auc_combined} where we specifically target ECC protection on the sensitive signed bit and we discover that, over the full window, our model still outperforms the targeted bit protection. This is because the leftover bits still add up since even if bit 7 is the largest single step, it is not the total error. Bits $0$--$5$ are $12.8\%$ of pooled SBU damage and bit~$6$ alone is $31.6\%$, using the aggregation above. SECDED(b7) leaves those unprotected. So at low $I$ enough lower-order bits are also flipped that the unprotected remainder dominates. This argument can still appear contradictory as it appears from Figures~\ref{fig:bitpos}(b) and \ref{fig:bitpos}(d) that the projection healing mechanism is predominantly sensitive to perturbations in bit 7. The key here is understanding this does \textit{not} make the mechanism identical to SECDED (b7). This is because projection healing \textit{can} correct all bits at once, while ECC is restricted to just 1 bit to correct. So under high multiplicity, bit 7 can often be flipped, which triggers both mechanisms (ECC and projection heal) but the damage is often not just bit 7 but multiple. Both methods equally detect the syndrome, but only the projection heal can apply the actual remedy. Hence we see performance improvements over SECDED (b7) at high intensity $I$ in Table~\ref{tab:rel_auc_combined}.

\subsection{Overhead}
\label{sec:compute_runtime}

Projection Healing adds an activation heal on every inference and a weight heal only at scheduled writeback. If the weight image is scrubbed once every $Q$ inference iterations, its projection cost is amortized by $Q$. We use $Q=256$ as an illustrative accounting cadence. On the square $N=128$, $k=32$ reference, the resulting average is $2.5\times$ baseline multiply-accumulates at batch size $1$ and $2.002\times$ at batch size $256$, reported as $2.5\times$ and $2.0\times$ in Table~\ref{tab:overhead}. The persistent model remains $1.0\times$ INT8 with no long-lived FP32 replica; a transient $4\times$ FP32 buffer exists only during each scrub. Appendix~\ref{app:overhead} gives the closed form.

\begin{table}[h]
\centering
\caption{Square-layer reference ($N{=}128$, $k{=}32$), with both projection passes always executed. Healing weights on every iteration costs $1+4k/N+4k/n_{\mathrm{mb}}$. Periodic INT8 writeback every $Q=256$ iterations amortizes the weight term to $1+4k/N+4k/(Qn_{\mathrm{mb}})$. ``ROM'' counts persistent model storage; periodic writeback uses a transient FP32 buffer during the scrub but retains no FP32 weight replica. Requantization, checksum, memory-traffic, and atomic-write costs are excluded.}
\label{tab:overhead}
\setlength{\tabcolsep}{3.5pt}
\begin{tabular}{@{}lccc@{}}
\hline
 &  & \multicolumn{2}{c}{FLOPs} \\
\cline{3-4}
Method & Persistent ROM & $n_{\mathrm{mb}}{=}256$ & $n_{\mathrm{mb}}{=}1$ \\
\hline
Unprotected & $1.0\times$ & $1.0\times$ & $1.0\times$ \\
Hardware TMR & $3.0\times$ & $3.0\times$ & $3.0\times$ \\
SECDED+intlv (72,64) & $1.125\times$ & ${\sim}1.0\times$ & ${\sim}1.0\times$ \\
SECDED(b7) UEP & $1.02\times$ & ${\sim}1.0\times$ & ${\sim}1.0\times$ \\
Clip.\ /\ Diehard & $1.0\times$ & ${\sim}1.0\times$ & ${\sim}1.0\times$ \\
Ours (heal every pass) & $\mathbf{1.0\times}$ & $\mathbf{2.5\times}$ & $\mathbf{{\sim}130\times}$ \\
Ours (writeback, $Q=256$) & $1.0\times$ & $\mathbf{2.0\times}$ & $\mathbf{2.5\times}$ \\
\hline
\end{tabular}
\end{table}

\section{Discussion}
\subsection{Comparisons}
We see across multiple datasets that projection healing provides measurable and significant improvement in reliability against MBU threats on static weights at high intensity regimes. We demonstrate this by isolating the mechanism and found that models receive a \ifsdsscolor\sdssnew{${\sim}23\%$}\else${\sim}22\%$\fi\ reduction in the native-unit excess integral of Table~\ref{tab:heal_contribution} in absolute terms. Across all datasets the method extends the operational scrub interval by a median of $2.04\times$ at a $+5\%$ RMSE degradation and $1.96\times$ at $+50\%$, compared to \ifsdsscolor\sdssnew{$1.62\times$ and $1.07\times$}\else$1.93\times$ and $1.15\times$\fi\ for classical Single Error Correction, Double Error Detection (SECDED). We further demonstrated that although significant bits are the primary damage contributor, strategically correcting those bits (SECDED (b7)) underperforms at high intensity regimes as multiple bits still remain uncorrected. 

We also demonstrated when compared to other models designed specifically for radiation-hardening techniques like DieHard and RedNet that those methods also fall short. In the wide window, Table~\ref{tab:rel_auc_combined} shows mean relative excess AUC near $0.9$ for RedNet and Diehard and \ifsdsscolor\sdssnew{$0.71$ (median $0.72$)}\else$0.70$ (median $0.71$)\fi\ for our method. That is further protection beyond those proxy baselines, whose main inference-time mechanism is activation bounding.

\subsection{Operational Environments}
\label{sec:overhead_summary}
We seek to characterize these improvements in real operational environments in order to translate measured gains to what might be theoretically expected on the field. Firstly, we know that $I$ can be interpreted as accumulated corruption until the weight image is scrubbed, thus the real practical payoff made possible is a longer window before a golden-copy reload is required.  We seek to produce an \emph{illustrative} conversion of performance into real calendar time in various settings.

Firstly, we model $I$ as a cumulative exposure with standard cross-section $I \approx \phi\,\sigma_{\mathrm{bit}}\,T$ with flux $\phi$, per-bit cross-section $\sigma_{\mathrm{bit}}$, and exposure $T$. For COTS sub-100~nm CMOS SRAM, $\sigma_{\mathrm{bit}}$ to high-energy hadrons is of order $10^{-14}$~cm$^2$/bit, with node-dependent scatter of roughly an order of magnitude \cite{ibe2010impact, dodd2003basic}. Mean multiplicity cancels between $I$ in Eq.~(\ref{eq:intensity}) and an empirical per-bit upset rate $\phi\sigma_{\mathrm{bit}}$ only when that rate already folds in $\mathbb{E}[m]$. 

In low Earth orbit, at ${\sim}10^{-6}$ upsets per bit per day \cite{falguere2002flight}, the CHF own-baseline $+5\%$ intensities are $I=6.32\times 10^{-4}$ ($1.7$ years) and $I=2.18\times 10^{-3}$ ($6.0$ years). Those times measure drift from each model's own clean RMSE. CHF already carries an $+11.8\%$ alignment tax, so a $+5\%$ budget measured from the unprotected floor is exhausted at $I=0$.

\textbf{Commercial avionics (12~km).} At ${\sim}4\times10^{-11}$ upsets per bit per hour \cite{normand1996single}, a $10^{4}$-hour lifetime accumulates only $I=4\times10^{-7}$, orders of magnitude below where even unprotected baselines degrade measurably, so the defense is unmotivated for benign aerospace and terrestrial use.

\textbf{Reactor in-core instrumentation.} Fluxes of $10^{13}$--$10^{14}$~n/cm$^2$/s \cite{kang1995design} give $dI/dt\approx 10^{-1}$~s$^{-1}$, so a parameter bank reaches $I=10^{-1}$ in one second and $I=0.3$ in three.

\subsection{Scope and Limits}

Despite the success presented, there are also clear limitations where this method proves to be neutral compared to other methods specifically at the low-intensity regimes where SBU dominates. We have shown that in lower intensity regimes, classical error-correction codes work in protecting these models, like with SECDED, when dominant errors are introduced by SBU rather than MBU. MBU becomes an issue at high intensity when the $\sim 33\%$ tail of MBU becomes problematic for single-bit correction to catch up.  

Furthermore, this method only makes operational sense in engineering regimes where specific trade-offs deem it necessary. At the illustrative $Q=256$ cadence, average square-layer arithmetic is $2.5\times$ at batch size $1$ and ${\sim}2.0\times$ at batch size $256$. On the six published widths the analytic single-sample range is $2.46\times$--$3.44\times$. The persistent footprint stays at $1.0\times$ INT8, but the scrub still needs transient workspace and reliable atomic writes. The cadence is not a measured radiation guarantee and must be tuned to the deployment upset rate. This could prove attractive in certain sectors where lead times are high, specifically given that advanced chip manufacturing is becoming a multi-year endeavor.

There also remain architectural limitations as well. We know that alignment with the pre-determined subspace comes at a cost in absolute model performance. This method also only makes sense if one's science case can tolerate a median $+0.3\%$ and mean ${\sim}2\%$ clean-RMSE tax, with CHF at $+11.8\pm 5.8\%$ as the worst case, for added $\times 2$ survivability over the duration of its deployment time. Furthermore, the healing mechanism only works when FAT has aligned the subspace (the no-weight-align arm in Section~\ref{sec:single_comp_rem} and Table~\ref{tab:ablation_fractional_auc}, not the frozen-checkpoint toggle in Section~\ref{sec:heal_attribution}). This means pre-trained models cannot be post-corrected by updating their gradients to realign their weights to the predetermined subspace. This makes the barrier to entry larger than passive techniques like simple error correction.

\section{Conclusion}
In this work we introduce Projection Healing, a software mechanism that corrects corrupted INT8 weights and pre-activations at runtime on unmodified commodity hardware. The mechanism works by projecting corrupted vectors into a protected low-rank subspace made possible by routing the network through this subspace via FAT-aligned training regularization. With periodic INT8 writeback every $Q=256$ iterations, the analytic square-layer average is $2.5\times$ baseline multiply-accumulates at batch size $1$ and ${\sim}2.0\times$ at batch size $256$, while retaining no long-lived FP32 weight replica (Appendix~\ref{app:overhead}). The cadence and repeated writeback behavior remain deployment parameters rather than evaluated radiation guarantees. We successfully demonstrated across six INT8 regression benchmarks that our method has the lowest relative excess AUC among the defenses we evaluated, which exclude in-place zero-space ECC and standalone unaligned weight-bit-flip FAT (Table~\ref{tab:rel_auc_combined}) which then extends the median scrub interval over unprotected storage at every tolerance we tested (Table~\ref{tab:breakdown_full}). Further, isolated on a frozen FAT checkpoint, the heal still contributes an additive gain (Section~\ref{sec:heal_attribution}). Replacing the projection with a MAD despike under the same alignment is worse (Table~\ref{tab:arch_retrain}). ECC remains competitive at low intensities, before multi-bit bursts saturate it. We believe our approach can sit alongside selective hardware protection on commodity hardware, where full weight TMR proves too expensive.

\appendix

\section{Compute Overhead}
\label{app:overhead}
For a square layer $W\in\mathbb{R}^{N\times N}$ the unprotected matmul costs $n_{\mathrm{mb}}N^{2}$ multiply-accumulates. Each robust heal is two projections, $4Nk$ multiply-accumulates, plus an $O(N)$ gate. Activation healing runs every iteration, while weight healing and INT8 writeback run once every $Q$ iterations. Always executing both projection passes gives the average
\begin{equation}
\mathrm{Ratio}_{Q}=1+\frac{4k}{N}+\frac{4k}{Q\,n_{\mathrm{mb}}}.
\end{equation}
At $N=128$, $k=32$, and $Q=256$, this is $2.002\times$ at $n_{\mathrm{mb}}=256$ and $2.5\times$ at $n_{\mathrm{mb}}=1$. Healing weights every iteration is the special case $Q=1$, giving $2.5\times$ and ${\sim}130\times$, respectively. On the published shapes, using $k=\min(32,N-1)$ and no activation heal on the scalar readout, periodic $Q=256$ writeback gives single-sample ratios $3.44\times$ for CalHousing and $2.46\times$--$2.50\times$ for the five wider models; at $n_{\mathrm{mb}}=256$ the ratios are $2.96\times$ and $1.97\times$--$2.01\times$. These are analytic multiply-accumulate ratios. They omit quantization, checksum, memory traffic, and transactional-write costs. The reported fault campaign did not run repeated writeback cycles, so it does not establish that $Q=256$ is an optimal or safe radiation cadence. The DCT basis is re-derivable and is omitted from the $1.0\times$ persistent-ROM column.

\section{Model Architectures and Capacities}
\label{app:architectures}
All six benchmarks use the same three-hidden-layer template of Section~\ref{sec:topological_healing}: $\mathrm{fc}_1$--$\mathrm{fc}_3$ with LeakyReLU, a linear readout, and INT8 post-training quantization. Widths land near two capacity bands (${\sim}18$\,k parameters for CalHousing, ${\sim}68$\,k for the rest). Table~\ref{tab:model_architectures} lists exact shapes, parameter counts, and INT8 footprints. Baselines are evaluated at the same widths so $I$ compares equal SRAM cross-sections within a dataset.

\input{tables/model_architectures}

\section{MBU Multiplicity Model}
\label{app:multiplicity}
\begin{figure}[t]
\centering
\includegraphics[width=0.8\textwidth]{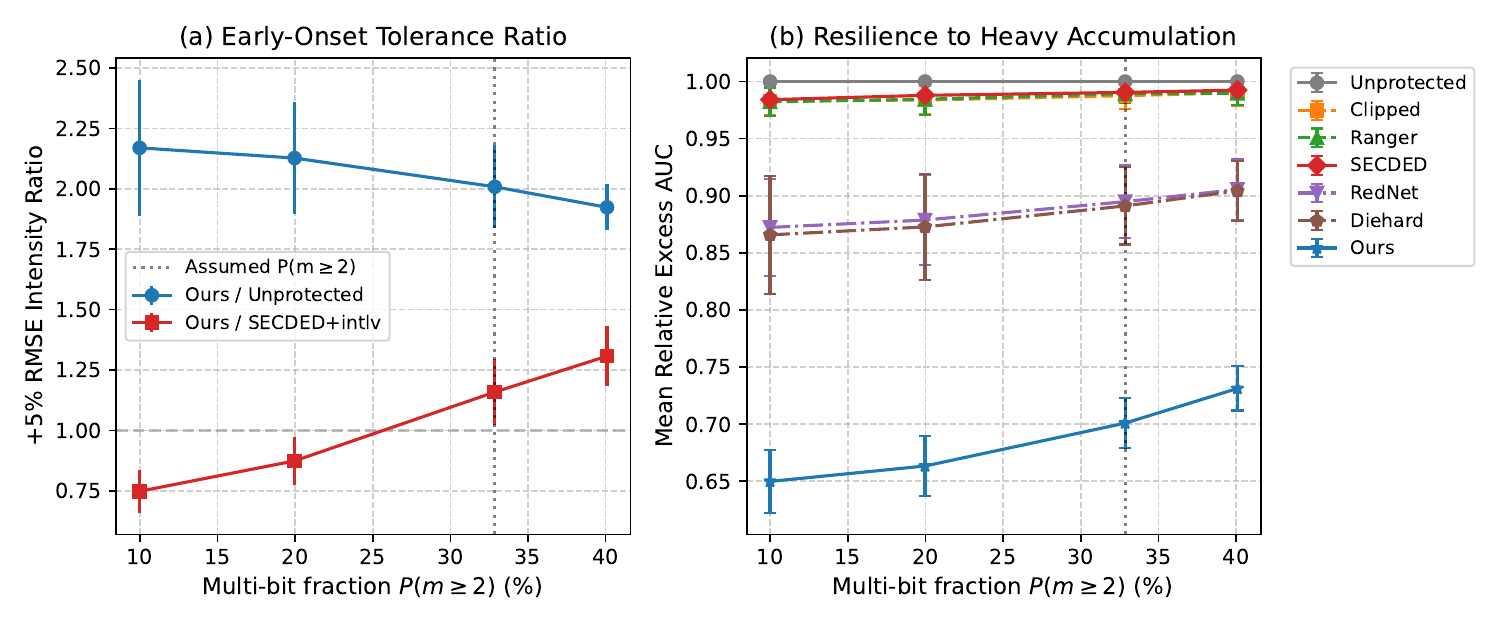}
\caption{Sensitivity to the assumed multiplicity spectrum. Both axes use $P(m\ge 2)$. (a)~Measured sensitivity of $+5\%$ RMSE, derived from Table~\ref{tab:breakdown_full}. (b)~Sensitivity of the six-dataset mean relative excess AUC over the wide window $I\in[10^{-4},1]$.}
\label{fig:alpha_sensitivity}
\end{figure}

Every fault in this paper draws its burst width $m$ from a single discrete power-law distribution $P(m)$ in Eq.~(\ref{eq:multiplicity}).  To pick the parameter we wish to strike, we uniformly sample a bit address in the flattened parameter store and flip $m$ contiguous bits with set parameters $\alpha=2.2$, $m_{\max}=256$ with some normalization factor $Z$.  
\begin{equation}
\begin{aligned}
P(m) &= m^{-\alpha}/Z,\\
m&\in\{1,\dots,m_{\max}\},\quad
Z=\sum_{m=1}^{m_{\max}} m^{-\alpha}.
\end{aligned}
\label{eq:multiplicity}
\end{equation}
\input{tables/mbu_multiplicity}
\textbf{Physical motivation and parameter choice.} Multiplicity in irradiated SRAM is a random variable: one ionizing track may upset a single cell or a cluster of adjacent cells, and measured cluster-size distributions are consistently dominated by single-bit events with a long tail that grows as feature size shrinks and as incidence approaches grazing \cite{black2008characterizing, dodd2003basic, baumann2005radiation, ibe2010impact}. We chose a truncated discrete power law since it is the simplest two-parameter family capturing a SBU-dominated distribution with a heavy tail. Given $m_{\max}$ cutoff $m_{\max}=256$ bits ($32$ INT8 words) Eq.~\ref{eq:multiplicity} places $\alpha$ in one-to-one correspondence with the multi-bit fraction $P(m\ge 2)$, so fixing one determines the other. We set $P(m\ge 2)=0.33$ and derive $\alpha=2.20$. This fraction is the same Table~II checkerboard event share cited in Section~\ref{sec:mbu_sim}, $29.1\%$ at $45$~nm and $38.5\%$ at $32$~nm \cite{ibe2010impact}, not a Table~III citation and not the odds $\mathrm{MCU}/\mathrm{SEU}$. Table~\ref{tab:mbu_multiplicity} separates strike fractions from flipped-bit shares. $P(m=1)=67.1\%$ of strikes, while those strikes are only $25.4\%$ of flipped bits. To further motivate our choice in distribution parameter we disentangle injection model impact by studying $\alpha$ sensitivity in a sweep documented in Figure~\ref{fig:alpha_sensitivity}. Here we re-run the evaluation sweep on the frozen eight-seed checkpoints at five alternative $(\alpha,m_{\max})$ settings spanning $P(m\ge 2)\in[10\%,40\%]$ and a $32\times$ $m_{max}$ range in cutoff, to separate which comparisons depend on the multiplicity spectrum. We report early-onset scrub-interval ratios at the $+5\%$ median-RMSE threshold of Table~\ref{tab:breakdown_full}, rather than solely wide-window relative excess AUC: in the saturated $I\in[10^{-4},1]$ integral SECDED sits at $0.98$--$0.99$ with almost no headroom, so spectrum dependence of the ECC comparison is invisible there.

The unprotected comparison is stable: at $+5\%$ rarely changing. The SECDED comparison is the sensitive one. At $+5\%$, Ours/SECDED rises from $0.75\times$ at a $10\%$ multi-bit fraction (SECDED ahead) through $1.16\times$ at the assumed $33\%$ to $1.31\times$ at $40\%$. The advantage over ECC therefore narrows as the environment becomes more single-bit dominated and reverses toward the pure-SBU limit, as Section~\ref{app:multiplicity} argued. The $+10\%$ column tracks the same monotone trend. However the $m_{max}$ knob is nearly inert: a $32\times$ sweep in $m_{\max}$ at fixed $\alpha=2.20$ keeps the $+5\%$ Ours/SECDED ratio inside $1.12\times$--$1.16\times$. This is to conclude that the assumed operating point does not dramatically benefit our work such that the narrative could have inverted under changes to the modeled multiplicity distribution. 

\section{Gate Threshold Sweep}
\label{app:gate_sweep}
We wanted to experiment how sensitive the gate threshold is. In Table~\ref{tab:gate_threshold_sweep} it shows a one-dimensional sweep of the weight-gate center $T$ at fixed $\tau_{\mathrm{flag}}=0.18$. Here bit~$6$ and bit~$7$ are mean row-detection excess and FPR is the zero-fault per-row fire rate. The cost is the clean-RMSE penalty relative to the heal-off and Rel.\ AUC is the wide-window excess integral divided by the $T=0.35$ value. We see that by lowering $T$ we can get more detection on these bit 6 faults but performance doesn't improve as this allows more false positives to sweep in and harm the overall score (Table~\ref{tab:gate_threshold_sweep}). 

\input{tables/gate_threshold_sweep}

\section{Surrogate vs.\ True INT8 Bit-Flip FAT}
\label{app:bitflip_fat}
Section~\ref{sec:fault_aware_training} trains under surrogates, instead of direct INT8 bit flips. Here we study the differences this may or may not create. We compare the surrogate FAT against an exact-fault-model variant. To implement the exact bit flips, we first draw INT8 spatial MBUs with exactly the evaluation geometry as our benchmark, and then pass through a straight-through estimator \cite{bengio2013estimating} $w\mapsto w+(\tilde{w}-w)_{\mathrm{detach}}$ so gradients still reach the float masters. In this experiment, every arm is trained on five seeds, paired on data split and evaluation fault stream, and swept at $200$ realizations per intensity knot. We report the nearest available curriculum severity without interpolating between discrete bands. As noted in Section~\ref{sec:leakage}, the aggregate performance difference at this nearest severity is only $3.3\%$ (Table~\ref{tab:bitflip_fat}), with the direction of the small residual varying by dataset. 
\input{tables/bitflip_fat}

\section{Architectural Validations and Sensitivity}
\label{app:arch_validations}
\subsection{fc$_1$ Heal No-Op}
The first-layer weight heal never fires on these networks. Their fan-ins are only $N\in[5,21]$, so a rank-$32$ DCT keeps every direction except one. That leftover direction spreads a fault almost evenly across the row. The gate requires a sharp peak, $\max(\mathbf{s})>2\,\mathrm{mean}(\mathbf{s})$, which this even spread cannot produce, so the row is never flagged. Turning the $\mathrm{fc}_1$ weight heal on or off is therefore bit-identical. A different basis that discarded a spikier direction could still trigger the gate, and a wider input layer would retain only $32$ of $N$ directions, so this no-op is specific to the DCT on these narrow layers. The layer is defended by the activation heal on its output, and retaining the call is harmless.

\subsection{Memory-Layout Sensitivity}
\label{app:interleave}
In our work we assumed a row-major fan-in addressing, under which a burst walks the fan-in of a single healed row. However other arrangements exist, like column-major addressing, which instead strides across rows at a fixed fan-in index, spreading damage over many healed rows at once. Table~\ref{tab:interleave} reports the ratio of column-major to row-major wide-window excess AUC under fixed burst widths of $3$, $24$, and $64$ bits ($1$, $3$, and $8$ INT8 words). \ifsdsscolor At one word the mean ratio is indistinguishable ($1.00$), as expected when the layouts differ only by a $1$--$2$ bit spill. Wider bursts leave the six-dataset mean near parity (\sdssnew{$1.01$} at three words, \sdssnew{$1.00$} at eight). JetPt carries the remaining degradation, SDSS stays near parity, and CMAPSS still improves.\else At one word the mean ratio is indistinguishable ($1.00$), as expected when the layouts differ only by a $1$--$2$ bit spill. Wider bursts do move the mean ($1.05$ at three words, $1.03$ at eight), with the largest degradation on SDSS and JetPt and a mild improvement on CMAPSS. The effect is modest on average (${<}5\%$), so row-major remains the favorable case rather than a neutral one, but layout is no longer a null result once bursts span multiple healed rows.\fi A further caveat is that physical and logical adjacency differ: production SRAMs bit-interleave and scramble addresses so that a physically contiguous multi-cell upset lands in several logical words \cite{ibe2010impact, baumann2005radiation}. Our injector draws bursts contiguous in the flattened logical address space, which is pessimistic for the interleaved SECDED baseline (it defeats the interleave more often than physical clustering would). Table~\ref{tab:interleave} shows the opposite direction for Projection Healing: a ratio above $1$ means column-major striding raises wide-window excess AUC relative to row-major. Spreading a burst across rows can drop each row below $(\tau_{\mathrm{flag}},T)$, so scrambling would be expected to help SECDED and slightly hurt the row-local heal. ``Interleaving'' elsewhere in the paper means SECDED word interleaving, not this tensor layout.

\input{tables/interleave}

\section{Sensitivity to the Green Rank Subspace Dimension}
\label{app:k_sweep}
One main architectural fix is the Green rank at $K=32$ on all hidden layers. While this isolates the architectural parameters, Projection Healing is a relative mechanism: the robust reconstruction depends on the ratio of the retained subspace to the full dimension ($k/N$). As noted in Appendix~\ref{app:arch_validations}, applying $K=32$ to a network with $N\le 32$ forces the basis to clamp to rank $k=N-1$, spanning nearly the entire feature space and making the gate structurally inert. 

To determine this sensitivity we evaluate a down-scaled version of the CalHousing benchmark with hidden dimensions $8\to64\to64\to48\to1$.  Table~\ref{tab:k_sweep} shows the results aggregated over an eight-seed campaign under identical spatial MBU injection. Projection Healing does not experience a brittle collapse as the subspace is restricted. Instead, resilience relative to the unprotected baseline falls as the subspace shrinks. A very tight bottleneck of $K=4$ introduces a minor alignment tax on the clean RMSE ($0.71$ vs.\ $0.53$) but still removes $30\%$ of the excess AUC compared to the unprotected network. The optimal configuration for this capacity is $K=16$ ($k/N \in [0.25, 0.33]$), which minimizes the relative damage ($0.54$ excess AUC) with minimal impact on clean convergence.

\input{tables/k_sweep_sensitivity}

\section{Baseline Implementations}
\label{app:baselines}
To ensure reproducibility and fair comparison in our simulated PyTorch environment, we specify the precise algorithmic implementations of each competitor arm evaluated in Section 4:

\begin{itemize}
    \item \textbf{Clipped:} Both weights and activations are clamped to the maximum dynamic ranges observed during a clean training pass. The clipping thresholds are static and applied per-layer.
    \item \textbf{Ranger \cite{chen2020ranger}:} Applies selective state-restriction to bound values based on training statistics. In our INT8 quantized environment, the bounds track the native quantization range, resulting in performance equivalent to the Clipped baseline.
    \item \textbf{SECDED+interleaving:} A standard $(72, 64)$ single-error-correction double-error-detection code in the Hamming/Hsiao family \cite{hamming1950error, hsiao1970class} is applied over the weight memory (a $1.125\times$ storage overhead). To mitigate multi-bit bursts (MBUs), it is augmented with a physical interleaving depth of $4$, dispersing adjacent spatial faults across different logical words, thus protecting a contiguous spatial span of $256$ bits ($32$ bytes). Check bits are modeled as protected and do not take strikes. The implementation models miscorrection: odd-weight errors $\ge 3$ alias to valid single-error syndromes, causing the decoder to flip an innocent bit and increase total damage \cite{chen1984error}. 
    \item \textbf{SECDED(b7) / unequal-error-protection:} The same $(72,64)$ SECDED decoder and interleave depth, but applied only to the gathered INT8 sign bits (bit~$7$) of the weight store. Sixty-four sign bits form one ECC data word, so the check-bit cost is $8/512=1.6\%$ of payload storage, one-eighth the full-SECDED overhead, while targeting the bit position that Figure~\ref{fig:bitpos} attributes the largest share of relative damage to. Lower-order bits are left unprotected. Miscorrection is modeled identically to the original with the caveat that an aliased flip now lands on a sign bit by construction, whereas a full-word miscorrection hits bit~$7$ only about one-eighth of the time.
    \item \textbf{RedNet proxy (LogClip):} each hidden layer is a linear map followed by LogClip, $\log(1+\mathrm{ReLU}(x))$ zeroed above $\theta=6$. There is no confidence head and no early exit. Scalar regression has no softmax probability to threshold. The original method \cite{ma2024rednet} also uses layer-sensitivity-aware spatial protection, which this proxy does not implement.
    \item \textbf{Diehard proxy:} uses ReLU6 hidden units and the spatial-MBU training injector. It does not clamp activations to $[0,1]$ and does not add a heavy $L_2$ penalty beyond the campaign weight decay.
    \item \textbf{Not evaluated.} In-place zero-space ECC \cite{guan2019inplace} trains weights so check bits occupy non-informative sign-extension slots and then protects those words with SEC-DED at no extra persistent ROM. We expect its single-bit window to resemble evaluated SECDED and its multi-bit ceiling to resemble SECDED once bursts exceed one correction. We do not report a six-dataset number for it. Standalone unaligned weight-bit-flip FAT \cite{stutz2021bit} is also absent. \ifsdsscolor The unaligned column of Table~\ref{tab:heal_contribution} removes subspace alignment and both runtime heals and keeps the surrogate curriculum.\fi\ Appendix~\ref{app:bitflip_fat} changes only the training injector on the full aligned, healed stack.
\end{itemize}

\data{The code and data supporting this study are not public at submission. The training code, fault-injection harness, and benchmark preprocessing scripts will be released, with the repository and license identified, when the paper has been accepted.}

\ack{P. Xiangyuan Ma thanks Jingxun Liang and Renfei Zhou for helpful discussions at the early stages of the project, and Joshua S. Bloom thanks Benjamin P. Nachman for useful comments at the later stages of the work. P. Xiangyuan Ma also thanks Tatsu K. for some early encouraging discussions regarding this project over bouldering at the gym while struggling to climb a stubborn V5 together. During the preparation of this manuscript, the authors used GPT-5.6 Sol, DeepSeek V4, Claude Opus 5.5, GLM-5.3, Kimi K3, and Gemini 3.1 Pro in code development and in writing and proof-editing. The authors reviewed and revised the assisted text and code and take responsibility for the content. The research directions, the experimental design, the evaluations, and the interpretation of the results are the work of the authors.}

\conflict{A provisional patent application regarding the methods in this manuscript has been filed by the University of California.}

\funding{P. Xiangyuan Ma acknowledges support from the Schmidt Science Foundation. We also acknowledge support from the Google Cloud Research Credits Program.}

\bibliographystyle{unsrt}
\bibliography{references}

\end{document}

%% file: sdss_color_switch.tex
% Main text: color-feature SDSS, including the Table 3 unaligned-FAT columns.
% Old text: published raw-magnitude SDSS, with no unaligned column.
% Restore the old text by changing the next line to \sdsscolorfalse.
\newif\ifsdsscolor
\sdsscolortrue
\newcommand{\sdssnew}[1]{#1}

%% file: tables/rel_auc_combined.tex
\begin{table}[t]
\centering
\caption{Normalized relative excess AUC, $\int\max(y-y^\star,0)\,d\log_{10}I$ divided by the unprotected integral. Sampled knots begin at $5\times 10^{-4}$. (a)~Early window through $10^{-2}$. (b)~Wide window through $1$. Lower is better. Entries are mean$\pm$std across eight seeds. The early-window mean is unstable where the unprotected denominator is near zero; use the median row there.}
\label{tab:rel_auc_combined}
\scriptsize
\setlength{\tabcolsep}{4pt}
\textbf{(a) Operational window} ($I \in [10^{-4}, 10^{-2}]$)\\[0.4em]
\begin{tabular}{lccccccc}
\hline
Dataset & Clipped & Ranger & RedNet & Diehard & SECDED & SECDED(b7) & \textbf{Ours} \\
\hline
CalHousing & $0.99 \pm 0.05$ & $1.03 \pm 0.11$ & $0.92 \pm 0.17$ & $0.60 \pm 0.15$ & $\mathbf{0.46 \pm 0.04}$ & $0.66 \pm 0.02$ & $0.47 \pm 0.15$ \\
Friedman & $1.00 \pm 0.00$ & $1.00 \pm 0.00$ & $1.38 \pm 0.30$ & $0.79 \pm 0.21$ & $\mathbf{0.53 \pm 0.08}$ & $0.56 \pm 0.07$ & $0.60 \pm 0.30$ \\
\ifsdsscolor
SDSS & \sdssnew{$1.00 \pm 0.29$} & \sdssnew{$1.03 \pm 0.09$} & \sdssnew{$1.20 \pm 0.55$} & \sdssnew{$0.96 \pm 0.41$} & \sdssnew{$\mathbf{0.44 \pm 0.15}$} & \sdssnew{$0.64 \pm 0.11$} & \sdssnew{$1.13 \pm 0.61$} \\
\else
SDSS & $1.02 \pm 0.05$ & $1.00 \pm 0.01$ & $0.97 \pm 0.21$ & $0.58 \pm 0.10$ & $\mathbf{0.38 \pm 0.06}$ & $0.63 \pm 0.06$ & $1.10 \pm 0.22$ \\
\fi
JetPt & $1.00 \pm 0.00$ & $1.00 \pm 0.01$ & $2.72 \pm 1.90$ & $1.64 \pm 1.02$ & $\mathbf{0.55 \pm 0.11}$ & $0.57 \pm 0.11$ & $0.69 \pm 0.59$ \\
CMAPSS & $1.00 \pm 0.00$ & $1.00 \pm 0.00$ & $0.81 \pm 1.78$ & $3.44 \pm 6.17$ & $0.81 \pm 0.29$ & $\mathbf{0.81 \pm 0.30}$ & $2.96 \pm 4.61$ \\
CHF & $1.00 \pm 0.00$ & $1.00 \pm 0.00$ & $0.95 \pm 0.13$ & $0.75 \pm 0.08$ & $\mathbf{0.42 \pm 0.02}$ & $0.64 \pm 0.04$ & $0.82 \pm 0.14$ \\
\hline
\ifsdsscolor
\textit{Median} & 1.00 & 1.00 & \sdssnew{1.07} & \sdssnew{0.88} & \textbf{0.49} & 0.64 & 0.76 \\
\textit{Mean} & 1.00 & 1.01 & \sdssnew{1.33} & \sdssnew{1.36} & \sdssnew{\textbf{0.54}} & 0.65 & 1.11 \\
\else
\textit{Median} & 1.00 & 1.00 & 0.96 & 0.77 & \textbf{0.49} & 0.64 & 0.76 \\
\textit{Mean} & 1.00 & 1.01 & 1.29 & 1.30 & \textbf{0.53} & 0.65 & 1.11 \\
\fi
\hline
\end{tabular}

\vspace{0.9em}
\textbf{(b) Wide window} ($I \in [10^{-4}, 1]$)\\[0.4em]
\begin{tabular}{lccccccc}
\hline
Dataset & Clipped & Ranger & RedNet & Diehard & SECDED & SECDED(b7) & \textbf{Ours} \\
\hline
CalHousing & $0.94 \pm 0.06$ & $0.96 \pm 0.05$ & $0.78 \pm 0.12$ & $0.75 \pm 0.15$ & $0.98 \pm 0.01$ & $1.00 \pm 0.01$ & $\mathbf{0.58 \pm 0.10}$ \\
Friedman & $1.00 \pm 0.00$ & $1.00 \pm 0.00$ & $0.96 \pm 0.02$ & $0.98 \pm 0.02$ & $0.99 \pm 0.00$ & $1.00 \pm 0.00$ & $\mathbf{0.79 \pm 0.02}$ \\
\ifsdsscolor
SDSS & \sdssnew{$0.97 \pm 0.04$} & \sdssnew{$0.98 \pm 0.03$} & \sdssnew{$0.91 \pm 0.17$} & \sdssnew{$0.88 \pm 0.14$} & \sdssnew{$1.00 \pm 0.01$} & \sdssnew{$1.01 \pm 0.01$} & \sdssnew{$\mathbf{0.71 \pm 0.10}$} \\
\else
SDSS & $0.98 \pm 0.01$ & $0.98 \pm 0.02$ & $0.90 \pm 0.11$ & $0.74 \pm 0.10$ & $0.99 \pm 0.01$ & $1.00 \pm 0.01$ & $\mathbf{0.63 \pm 0.09}$ \\
\fi
JetPt & $1.00 \pm 0.00$ & $1.00 \pm 0.00$ & $0.95 \pm 0.02$ & $0.99 \pm 0.02$ & $1.00 \pm 0.00$ & $1.01 \pm 0.00$ & $\mathbf{0.78 \pm 0.02}$ \\
CMAPSS & $1.00 \pm 0.00$ & $1.00 \pm 0.00$ & $0.89 \pm 0.03$ & $1.00 \pm 0.02$ & $1.01 \pm 0.00$ & $1.01 \pm 0.01$ & $\mathbf{0.69 \pm 0.04}$ \\
CHF & $1.00 \pm 0.00$ & $0.99 \pm 0.01$ & $0.89 \pm 0.08$ & $0.88 \pm 0.07$ & $0.98 \pm 0.00$ & $1.00 \pm 0.00$ & $\mathbf{0.73 \pm 0.07}$ \\
\hline
\ifsdsscolor
\textit{Median} & 1.00 & 0.99 & 0.90 & 0.93 & 0.99 & 1.00 & \sdssnew{\textbf{0.72}} \\
\textit{Mean} & \sdssnew{0.98} & 0.99 & 0.90 & \sdssnew{0.91} & 0.99 & 1.00 & \sdssnew{\textbf{0.71}} \\
\else
\textit{Median} & 1.00 & 0.99 & 0.90 & 0.93 & 0.99 & 1.00 & \textbf{0.71} \\
\textit{Mean} & 0.99 & 0.99 & 0.90 & 0.89 & 0.99 & 1.00 & \textbf{0.70} \\
\fi
\hline
\end{tabular}
\end{table}

%% file: tables/breakdown_full.tex
\begin{table}[t]
\centering
\caption{Exposure intensity $I$ at which median test RMSE rises $+5\%$, $+10\%$, and $+50\%$ above each model's own clean baseline, not the shared unprotected floor. Bold indicates the largest $I$. The $+50\%$ column is $I_{\mathrm{break}}$, not untrained-network collapse.}
\label{tab:breakdown_full}
\scriptsize
\setlength{\tabcolsep}{3pt}
\resizebox{\textwidth}{!}{%
\begin{tabular}{l ccc ccc ccc}
\hline
& \multicolumn{3}{c}{\textbf{+5\% RMSE (Operational)}} & \multicolumn{3}{c}{\textbf{+10\% RMSE (Degraded)}} & \multicolumn{3}{c}{\textbf{+50\% RMSE (Breakdown)}} \\
\cline{2-4} \cline{5-7} \cline{8-10}
Dataset & Unprot. & SECDED & \textbf{Ours} & Unprot. & SECDED & \textbf{Ours} & Unprot. & SECDED & \textbf{Ours} \\
\hline
CalHousing & $2.38{\times}10^{-3}$ & $4.98{\times}10^{-3}$ & $\mathbf{5.31{\times}10^{-3}}$ & $4.19{\times}10^{-3}$ & $7.82{\times}10^{-3}$ & $\mathbf{9.49{\times}10^{-3}}$ & $2.22{\times}10^{-2}$ & $2.57{\times}10^{-2}$ & $\mathbf{5.68{\times}10^{-2}}$ \\
Friedman & $3.34{\times}10^{-3}$ & $5.94{\times}10^{-3}$ & $\mathbf{6.18{\times}10^{-3}}$ & $5.99{\times}10^{-3}$ & $9.30{\times}10^{-3}$ & $\mathbf{1.11{\times}10^{-2}}$ & $2.36{\times}10^{-2}$ & $2.72{\times}10^{-2}$ & $\mathbf{4.24{\times}10^{-2}}$ \\
\ifsdsscolor
SDSS & \sdssnew{$9.06{\times}10^{-3}$} & \sdssnew{$\mathbf{1.33{\times}10^{-2}}$} & \sdssnew{$1.33{\times}10^{-2}$} & \sdssnew{$1.51{\times}10^{-2}$} & \sdssnew{$1.98{\times}10^{-2}$} & \sdssnew{$\mathbf{2.27{\times}10^{-2}}$} & \sdssnew{$5.54{\times}10^{-2}$} & \sdssnew{$5.50{\times}10^{-2}$} & \sdssnew{$\mathbf{9.24{\times}10^{-2}}$} \\
\else
SDSS & $2.32{\times}10^{-3}$ & $\mathbf{5.03{\times}10^{-3}}$ & $3.13{\times}10^{-3}$ & $4.15{\times}10^{-3}$ & $\mathbf{7.71{\times}10^{-3}}$ & $5.55{\times}10^{-3}$ & $2.34{\times}10^{-2}$ & $2.67{\times}10^{-2}$ & $\mathbf{3.54{\times}10^{-2}}$ \\
\fi
JetPt & $1.81{\times}10^{-2}$ & $2.20{\times}10^{-2}$ & $\mathbf{3.18{\times}10^{-2}}$ & $2.85{\times}10^{-2}$ & $3.17{\times}10^{-2}$ & $\mathbf{5.00{\times}10^{-2}}$ & $9.99{\times}10^{-2}$ & $9.48{\times}10^{-2}$ & $\mathbf{1.66{\times}10^{-1}}$ \\
CMAPSS & $2.34{\times}10^{-2}$ & $2.68{\times}10^{-2}$ & $\mathbf{5.31{\times}10^{-2}}$ & $3.42{\times}10^{-2}$ & $3.62{\times}10^{-2}$ & $\mathbf{7.99{\times}10^{-2}}$ & $1.06{\times}10^{-1}$ & $1.01{\times}10^{-1}$ & $\mathbf{2.25{\times}10^{-1}}$ \\
CHF & $6.32{\times}10^{-4}$ & $1.90{\times}10^{-3}$ & $\mathbf{2.18{\times}10^{-3}}$ & $1.12{\times}10^{-3}$ & $2.93{\times}10^{-3}$ & $\mathbf{3.66{\times}10^{-3}}$ & $5.31{\times}10^{-3}$ & $9.14{\times}10^{-3}$ & $\mathbf{1.46{\times}10^{-2}}$ \\
\hline
\ifsdsscolor
\textit{Median Extension} & -- & \sdssnew{1.62$\times$} & \textbf{2.04$\times$} & -- & \sdssnew{1.43$\times$} & \textbf{2.06$\times$} & -- & \sdssnew{1.07$\times$} & \textbf{1.96$\times$} \\
\else
\textit{Median Extension} & -- & 1.93$\times$ & \textbf{2.04$\times$} & -- & 1.71$\times$ & \textbf{2.06$\times$} & -- & 1.15$\times$ & \textbf{1.96$\times$} \\
\fi
\hline
\end{tabular}}
\end{table}

%% file: tables/heal_contribution.tex
\begin{table}[t]
\centering
\ifsdsscolor
\caption{Native-unit excess $\int\max(y-y^\star,0)\,d\log_{10}I$ on $I\in[5\times 10^{-4},\,3\times 10^{-1}]$, mean$\pm$std across seeds. Not normalized by the unprotected integral. Gain is $(\mathrm{No\ heal}-\mathrm{Full})/\mathrm{No\ heal}$ from unrounded values. Unaligned is a from-scratch retrain with the alignment penalties and both runtime heals removed; the surrogate curriculum and range penalty stay. Vs.\ unaligned is $(\mathrm{Unaligned}-\mathrm{Full})/\mathrm{Unaligned}$. Clean $\Delta$ is the fault-free RMSE change from toggling the heal on one frozen aligned checkpoint. The SDSS row uses the colors $r$, $u{-}g$, $g{-}r$, $r{-}i$, and $i{-}z$.}
\label{tab:heal_contribution}
\setlength{\tabcolsep}{2.2pt}
\resizebox{\textwidth}{!}{%
\begin{tabular}{lcccccrrr}
\hline
Dataset & Full & No wt.\ heal & No act.\ heal & No heal & Unaligned & Gain & Vs.\ unaligned & Clean $\Delta$ \\
\hline
CalHousing & $0.40\pm0.05$ & $0.42\pm0.04$ & $0.46\pm0.06$ & $0.48\pm0.06$ & \sdssnew{$0.43\pm0.06$} & 15.9\% & \sdssnew{8.0\%} & +0.3\% \\
Friedman & $1.43\pm0.05$ & $1.78\pm0.06$ & $1.52\pm0.04$ & $1.86\pm0.05$ & \sdssnew{$1.65\pm0.06$} & 23.5\% & \sdssnew{14.0\%} & +1.0\% \\
SDSS & \sdssnew{$0.012\pm0.001$} & \sdssnew{$0.013\pm0.001$} & \sdssnew{$0.014\pm0.001$} & \sdssnew{$0.015\pm0.001$} & \sdssnew{$0.012\pm0.001$} & \sdssnew{22.3\%} & \sdssnew{$-$0.2\%} & \sdssnew{+0.4\%} \\
JetPt & $8.96\pm0.37$ & $10.9\pm0.3$ & $9.83\pm0.46$ & $11.7\pm0.4$ & \sdssnew{$11.6\pm0.8$} & 23.2\% & \sdssnew{23.1\%} & +0.1\% \\
CMAPSS & $4.03\pm0.59$ & $5.96\pm0.69$ & $4.18\pm0.62$ & $6.08\pm0.72$ & \sdssnew{$5.34\pm0.87$} & 33.7\% & \sdssnew{25.4\%} & $-$0.2\% \\
CHF & $641.9\pm64.2$ & $734.4\pm49.6$ & $748.7\pm103.3$ & $823.3\pm89.7$ & \sdssnew{$730\pm45$} & 22.0\% & \sdssnew{11.9\%} & +1.6\% \\
\hline
\textbf{Mean} & \multicolumn{5}{c}{} & \sdssnew{\textbf{23.4\%}} & \sdssnew{\textbf{13.7\%}} & \sdssnew{\textbf{+0.5\%}} \\
\hline
\end{tabular}}
\else
\caption{Native-unit excess $\int\max(y-y^\star,0)\,d\log_{10}I$ on $I\in[5\times 10^{-4},\,3\times 10^{-1}]$, mean$\pm$std across seeds. Not normalized by the unprotected integral. Gain is $(\mathrm{No\ heal}-\mathrm{Full})/\mathrm{No\ heal}$ from unrounded values. Clean $\Delta$ is the fault-free RMSE change from toggling the heal on one frozen FAT checkpoint.}
\label{tab:heal_contribution}
\setlength{\tabcolsep}{3.0pt}
\begin{tabular}{lccccrr}
\hline
Dataset & Full & No wt.\ heal & No act.\ heal & No heal & Gain & Clean $\Delta$ \\
\hline
CalHousing & $0.40\pm0.05$ & $0.42\pm0.04$ & $0.46\pm0.06$ & $0.48\pm0.06$ & 15.9\% & +0.3\% \\
Friedman & $1.43\pm0.05$ & $1.78\pm0.06$ & $1.52\pm0.04$ & $1.86\pm0.05$ & 23.5\% & +1.0\% \\
SDSS & $0.02\pm0.00$ & $0.02\pm0.00$ & $0.03\pm0.00$ & $0.03\pm0.00$ & 14.6\% & +1.8\% \\
JetPt & $8.96\pm0.37$ & $10.9\pm0.3$ & $9.83\pm0.46$ & $11.7\pm0.4$ & 23.2\% & +0.1\% \\
CMAPSS & $4.03\pm0.59$ & $5.96\pm0.69$ & $4.18\pm0.62$ & $6.08\pm0.72$ & 33.7\% & $-$0.2\% \\
CHF & $641.9\pm64.2$ & $734.4\pm49.6$ & $748.7\pm103.3$ & $823.3\pm89.7$ & 22.0\% & +1.6\% \\
\hline
\textbf{Mean} & \multicolumn{4}{c}{} & \textbf{22.2\%} & \textbf{+0.8\%} \\
\hline
\end{tabular}
\fi
\end{table}

%% file: tables/arch_retrain.tex
\begin{table}[!t]
\centering
\caption{Absolute integral $\int y\,d\log_{10}I$ in native target units. Parentheses are the ratio to the DCT column. Lower is better. CHF, not the other five datasets, is sensitive to the basis.}
\label{tab:arch_retrain}
\scriptsize
\setlength{\tabcolsep}{3.2pt}
\begin{tabular}{lccccccccc}
\hline
 & \multicolumn{3}{c}{Full (DCT)} & \multicolumn{3}{c}{Arbitrary basis} & \multicolumn{3}{c}{MAD despike} \\
\hline
CalHousing & \multicolumn{3}{c}{2.24}& \multicolumn{3}{c}{\textbf{2.22 (0.99)}}& \multicolumn{3}{c}{2.87 (1.28)}\\
Friedman & \multicolumn{3}{c}{5.38}& \multicolumn{3}{c}{\textbf{5.02 (0.93)}}& \multicolumn{3}{c}{10.05 (1.87)}\\
\ifsdsscolor
SDSS & \multicolumn{3}{c}{\textbf{0.10}}& \multicolumn{3}{c}{\sdssnew{0.11 (1.09)}}& \multicolumn{3}{c}{\sdssnew{0.12 (1.22)}}\\
\else
SDSS & \multicolumn{3}{c}{\textbf{0.10}}& \multicolumn{3}{c}{0.12 (1.19)}& \multicolumn{3}{c}{0.13 (1.32)}\\
\fi
JetPt & \multicolumn{3}{c}{101.4}& \multicolumn{3}{c}{\textbf{98.6 (0.97)}}& \multicolumn{3}{c}{120.3 (1.19)}\\
CMAPSS & \multicolumn{3}{c}{65.8}& \multicolumn{3}{c}{\textbf{64.5 (0.98)}}& \multicolumn{3}{c}{70.6 (1.07)}\\
CHF & \multicolumn{3}{c}{\textbf{1529}}& \multicolumn{3}{c}{2721 (1.78)}& \multicolumn{3}{c}{2843 (1.86)}\\
\hline
\end{tabular}
\end{table}

%% file: tables/model_architectures.tex
\begin{table}[t]
\centering
\caption{Per-dataset MLP shapes. INT8 bits equal parameter count times $8$, counting each bias element as one stored byte. Per-channel scales are extra protected metadata and are not in this count.}
\label{tab:model_architectures}
\scriptsize
\setlength{\tabcolsep}{3pt}
\begin{tabular}{@{}lccc@{}}
\hline
Dataset & Arch. & Params & INT8 bits \\
\hline
CalHousing & $8{\to}96{\to}96{\to}80{\to}1$ & 18\,017 & 144\,136 \\
Friedman & $10{\to}200{\to}200{\to}128{\to}1$ & 68\,257 & 546\,056 \\
SDSS & $5{\to}200{\to}200{\to}128{\to}1$ & 67\,257 & 538\,056 \\
JetPt & $13{\to}200{\to}200{\to}128{\to}1$ & 68\,857 & 550\,856 \\
CMAPSS & $21{\to}200{\to}200{\to}128{\to}1$ & 70\,457 & 563\,656 \\
CHF & $5{\to}200{\to}200{\to}128{\to}1$ & 67\,257 & 538\,056 \\
\hline
\end{tabular}
\end{table}

%% file: tables/mbu_multiplicity.tex
\begin{table}[t]
\centering
\caption{Properties of the truncated discrete power-law multiplicity $P(m)=m^{-\alpha}/Z$ on $m\in\{1,\dots,m_{\max}\}$ with $\alpha=2.2$, $m_{\max}=256$. Exact arithmetic, not measurements. The median of $m$ is $1$. Bit share is the fraction of flipped bits, not the fraction of strikes.}
\label{tab:mbu_multiplicity}
\setlength{\tabcolsep}{4pt}
\begin{tabular}{lcc}
\hline
Quantity & Value & Bit share \\
\hline
$Z=\sum_{m=1}^{256} m^{-2.2}$ & $1.4895$ & -- \\
$P(m=1)$ (SBU) & $67.1\%$ & $25.4\%$ \\
$P(m\ge 2)$ & $32.9\%$ & $74.6\%$ \\
$P(m\ge 8)$ & $4.90\%$ & $42.9\%$ \\
$P(m\ge 32)$ & $0.82\%$ & $21.8\%$ \\
$P(m\ge 128)$ & $0.09\%$ & $6.3\%$ \\
\hline
Mean multiplicity $\mathbb{E}[m]$ & $2.65$ & -- \\
Std.\ dev.\ $\sigma_m$ & $7.97$ & -- \\
Mean INT8 words spanned & $1.21$ & -- \\
\hline
\end{tabular}
\end{table}

%% file: tables/gate_threshold_sweep.tex
\begin{table}[t]
\centering
\caption{Weight-gate center sweep at $\tau_{\mathrm{flag}}=0.18$ on seed $0$. Bit columns are row-detection excess. FPR is the zero-fault row fire rate. Cost is the clean-RMSE penalty versus the heal-off end. Rel.\ AUC is the ratio to $T=0.35$. Dashes are ladder ends without that ratio. The dagger is the weight operating point, not the activation center $T=0.25$.}
\label{tab:gate_threshold_sweep}
\setlength{\tabcolsep}{2.0pt}
\footnotesize
\begin{tabular}{lccccc}
\hline
$T$ & Bit~6 & Bit~7 & FPR (\%) & Cost (\%) & Rel.\ AUC \\
\hline
0.04 & 0.27 & 0.58 & 24.7 & 23.9 & -- \\
0.08 & 0.27 & 0.58 & 24.7 & 13.6 & -- \\
0.12 & 0.27 & 0.58 & 24.7 & 8.6 & -- \\
0.14$^{\ast}$ & 0.27 & 0.58 & 24.7 & 6.6 & 1.10 \\
0.20 & 0.21 & 0.62 & 22.0 & 3.1 & 1.04 \\
0.27 & 0.09 & 0.71 & 15.0 & 1.3 & 1.01 \\
0.35$^{\dagger}$ & 0.04 & 0.77 & 9.1 & 0.5 & 1.00 \\
0.45 & 0.02 & 0.74 & 5.1 & 0.1 & 1.01 \\
0.57 & 0.01 & 0.62 & 2.6 & 0.0 & -- \\
\hline
\end{tabular}
\\[2pt]
\footnotesize $^{\dagger}$published operating point; $^{\ast}$gate center at which bit 6 first clears the syndrome threshold.
\end{table}

%% file: tables/bitflip_fat.tex
\begin{table}[t]
\centering
\ifsdsscolor
\caption{Full-stack curriculum ablation. Every arm still includes subspace alignment and eval-time Projection Healing. Entries are paired ratios of wide-window excess AUC to the published surrogate curriculum; below one favors that column's arm. BF (orig.) is the original bit-flip severity, about $6\times$ milder than the surrogate. BF (near) matches the surrogate on measured severity. Surr.+STE sends the surrogate masks through the bit-flip arm's straight-through estimator. A dagger marks a paired bootstrap interval that excludes parity. Parentheses on the geometric-mean row are bootstrap ranges. This is not a standalone unaligned FAT baseline. The SDSS entry in BF (orig.) uses the color features, and that column's geometric mean has no bootstrap range because BF (near) and Surr.+STE were not rerun.}
\else
\caption{Full-stack curriculum ablation. Every arm still includes subspace alignment and eval-time Projection Healing. Entries are paired ratios of wide-window excess AUC to the published surrogate curriculum; below one favors that column's arm. BF (orig.) is the original bit-flip severity, about $6\times$ milder than the surrogate. BF (near) matches the surrogate on measured severity. Surr.+STE sends the surrogate masks through the bit-flip arm's straight-through estimator. A dagger marks a paired bootstrap interval that excludes parity. Parentheses on the geometric-mean row are bootstrap ranges. This is not a standalone unaligned FAT baseline.}
\fi
\label{tab:bitflip_fat}
\footnotesize
\setlength{\tabcolsep}{2.0pt}

\begin{tabular}{lccc}
\hline
Dataset & BF (orig.) & BF (near) & Surr.+STE \\
\hline
CalHousing & 1.09$^{\dagger}$ & 0.88$^{\dagger}$ & 0.98 \\
Friedman & 1.07$^{\dagger}$ & 1.04$^{\dagger}$ & 1.05$^{\dagger}$ \\
\ifsdsscolor
SDSS & \sdssnew{0.87$^{\dagger}$} & 0.92$^{\dagger}$ & 0.97 \\
\else
SDSS & 1.12$^{\dagger}$ & 0.92$^{\dagger}$ & 0.97 \\
\fi
JetPt & 1.14$^{\dagger}$ & 0.89$^{\dagger}$ & 0.94$^{\dagger}$ \\
CMAPSS & 1.15$^{\dagger}$ & 1.06$^{\dagger}$ & 1.07$^{\dagger}$ \\
CHF & 1.07$^{\dagger}$ & 1.02$^{\dagger}$ & 1.02$^{\dagger}$ \\
\hline
\ifsdsscolor
Geo.\ mean & \sdssnew{\textbf{1.06}} & \textbf{0.97} & \textbf{1.00} \\
 &  & {\scriptsize(0.91--1.03)} & {\scriptsize(0.97--1.04)} \\
\else
Geo.\ mean & \textbf{1.10} & \textbf{0.97} & \textbf{1.00} \\
 & {\scriptsize(1.08--1.13)} & {\scriptsize(0.91--1.03)} & {\scriptsize(0.97--1.04)} \\
\fi
\hline
\end{tabular}%
\end{table}

%% file: tables/interleave.tex
\begin{table}[t]
\centering
\caption{Ratio of column-major to row-major wide-window excess AUC. A ratio above $1$ means the column-major layout degrades Projection Healing relative to row-major. This is tensor layout, not SECDED word interleaving.}
\label{tab:interleave}
\setlength{\tabcolsep}{4pt}
\begin{tabular}{lccc}
\hline
Dataset & $S_{\mathrm{block}}=3$ (1w) & $S_{\mathrm{block}}=24$ (3w) & $S_{\mathrm{block}}=64$ (8w) \\
\hline
CalHousing & 1.02 & 1.04 & 1.08 \\
Friedman & 0.99 & 1.04 & 1.01 \\
\ifsdsscolor
SDSS & \sdssnew{1.00} & \sdssnew{1.02} & \sdssnew{0.99} \\
\else
SDSS & 1.03 & 1.24 & 1.19 \\
\fi
JetPt & 1.04 & 1.06 & 1.16 \\
CMAPSS & 0.93 & 0.84 & 0.76 \\
CHF & 1.03 & 1.07 & 1.00 \\
\hline
\ifsdsscolor
\textbf{Mean} & \textbf{1.00} & \sdssnew{\textbf{1.01}} & \sdssnew{\textbf{1.00}} \\
\else
\textbf{Mean} & \textbf{1.00} & \textbf{1.05} & \textbf{1.03} \\
\fi
\hline
\end{tabular}
\end{table}

%% file: tables/k_sweep_sensitivity.tex
\begin{table}[h]
\centering
\caption{Sensitivity of Projection Healing to the Green rank $K$ on a narrow architecture. Column heads are abbreviated so the table fits the column; $k/N$ is the retained-rank ratio on each hidden layer.}
\label{tab:k_sweep}
\footnotesize
\begin{tabular}{lccc}
\hline
\textbf{Model} & $k/N$ & \textbf{Clean RMSE} & Rel.\ AUC \\
\hline
Unprotected  & --                      & $0.53 \pm 0.01$ & $1.00 \pm 0.00$ \\
$K=4$        & 0.06, 0.06, 0.08     & $0.71 \pm 0.05$ & $0.71 \pm 0.12$ \\
$K=8$        & 0.13, 0.13, 0.17     & $0.59 \pm 0.02$ & $0.56 \pm 0.09$ \\
$K=16$       & 0.25, 0.25, 0.33     & $0.55 \pm 0.01$ & $0.54 \pm 0.08$ \\
$K=32$       & 0.50, 0.50, 0.67     & $0.53 \pm 0.01$ & $0.59 \pm 0.10$ \\
\hline
\end{tabular}
\end{table}